\documentclass[fleqn,usenatbib]{mnras}

\usepackage{newtxtext,newtxmath}

\usepackage[T1]{fontenc}

\DeclareRobustCommand{\VAN}[3]{#2}
\let\VANthebibliography\thebibliography
\def\thebibliography{\DeclareRobustCommand{\VAN}[3]{##3}\VANthebibliography}

\usepackage{graphicx}	
\usepackage{amsmath}	
\usepackage{multirow}

\newcommand{\hb}{H$\beta$}
\newcommand{\hg}{H$\gamma$}

\newcommand{\mgii}{\ion{Mg}{ii}}
\newcommand{\nev}{[\ion{Ne}{v}]}
\newcommand{\oii}{[\ion{O}{ii}]}
\newcommand{\neiii}{[\ion{Ne}{iii}]}
\newcommand{\oiii}{[\ion{O}{iii}]}
\newcommand{\oiiid}{[\ion{O}{iii}]$\lambda\lambda 4959,5007$}
\newcommand{\feii}{\ion{Fe}{ii}}
\newcommand{\hei}{\ion{He}{i}}

\newcommand{\kms}{\textrm{km}\,\textrm{s}$^{-1}$}
\newcommand{\ergs}{\textrm{erg}\,\textrm{s}$^{-1}$}

\newcommand{\lsun}{L$_\odot$}
\newcommand{\msun}{M$_\odot$}

\title[Differences in the NLR of QSOs 1 and 2 -- I]{The differences in the Narrow Line Region of nearby QSOs 1 and 2 -- I: higher excitation and contribution of shocks in type 1's} 

\author[Hauschild-Roier et al.]{
Gabriel R. Hauschild-Roier,$^{1}$\thanks{E-mail: gabrielrhroier@gmail.com}
Thaisa Storchi-Bergmann$^{1}$
Rogério Riffel$^{1}$
and Vincenzo Mainieri$^{2}$
\\
$^{1}$Departamento de Astronomia, Universidade Federal do Rio Grande do Sul, Av. Bento Gonçalves 9500, 91501-970, Porto Alegre, RS, Brazil\\
$^{2}$European Southern Observatory, Karl-Schwarzschild-Straße 2, D-
85748 Garching bei Munchen, Germany
}

\date{Accepted XXX. Received YYY; in original form ZZZ}

\pubyear{\the\year{}}

\begin{document}
\label{firstpage}
\pagerange{\pageref{firstpage}--\pageref{lastpage}}
\maketitle

\begin{abstract}

We compare the excitation of the Narrow-Line Region (NLR) of type 1 and type 2 QSOs for redshifts $0.4 \le z \le 0.5$ via the analysis of their emission line properties in Sloan Digital Sky Survey (SDSS) near-UV/optical spectra. We fit the continuum and emission lines, using two kinematic components for \oiii$\lambda$5007 and \hb\ (narrow and broad) and a single component for the weaker lines. We find two main differences in the NLR excitation of type 1 and 2 QSOs:
(i) QSOs 2 have higher \oiii/\hb\ than QSOs 1 in both narrow and broad components; (ii) QSOs 1 present higher \nev, \neiii\ and \oiii$\lambda4363$ luminosities, higher \nev/\neiii\ and \neiii/\oii\ ratios and higher temperatures than  QSOs 2. These differences support more highly excited regions, higher temperature gas and prevalence of shocks in type 1 relative to type 2 QSOs. We suggest two possible scenarios: (i) type 1 QSOs are seen more pole-on, allowing the observation of more highly excited gas closer to the nucleus, supporting the Unified Model scenario; (ii) evolution from type 2 to type 1 QSOs, with highest excitation regions obscured in type 2's and cleared up in a ``blow-out phase". Support for the evolutionary scenario is given by the usually higher L\oiii\ in QSOs 2, in the sense that these sources host a more powerful AGN that, in its evolution, clears up the excess dust and gas to reveal a lower-luminosity but more highly excited type 1 AGN.

\end{abstract}

\begin{keywords}
galaxies: active --  quasars: emission lines -- quasars: general
\end{keywords}



\section{Introduction} \label{sec:intro}

Quasars are active galactic nuclei (AGN) and the most luminous objects in the universe. In their center a supermassive black hole (SMBH) is being fed by mass accretion \citep{kormendy_coevolution_2013} at the highest accretion rates, with bolometric luminosities $L_\mathrm{bol}\,>\,10^{46}$\ergs. Galaxy evolution models describe major episodes of SMBHs growth occurring in phases in which the AGN is obscured during the episodes that led to its triggering, e.g. via mass accretion to the nucleus that occurs, for example, in luminous or ultra-luminous infrared galaxies \citep[ULIRGS, e.g.][]{sanders_ultraluminous_1988, hopkins_black_2005, hopkins_quasar_2010, veilleux_spitzer_2009, simpson_alma_2014}. The end phase of this growth is a blowout phase that expels the gas and dust revealing the nuclear engine of the AGN -- the blue continuum emission from the accretion disk and the broad emission lines from clouds that surround the disk and form the Broad-Line Region (BLR).

The dichotomy between obscured and unobscured AGN is also postulated by the Unified Model, that attributes the observed differences between Type 1 (showing a blue continuum and visible BLR) and Type 2 (redder continuum and only narrow emission lines) AGN to orientation effects associated with an axisymmetric dusty torus surrounding the central engine \citep{antonucci_unified_1993, urry_unified_1995, nenkova_dust_2002, netzer_revisiting_2015, ramos_almeida_nuclear_2017, harrison_observational_2024}. Type 1 AGN offer direct views of the central engine and BLR while in Type 2s these regions are heavily obscured due to our nearly edge-on view of the torus/accretion disc geometry.

Although there are many evidences that support the Unified Model, in which the difference between type 1 and type 2 objects is only due to orientation -- e.g. the presence of BLR emission in polarized light in type 1 AGN \citep{antonucci_spectropolarimetry_1985, antonucci_unified_1993}, there are also evidences of intrinsic differences between the two types, which suggest that type 2 AGN is an obscured phase that evolves to type 1 AGN in an unobscured phase \citep[e.g.][]{sanders_ultraluminous_1988, veilleux_spitzer_2009, ramos_almeida_testing_2011, audibert_probing_2017}.


In the case of QSOs, recent studies of a nearby sample (z$\le$0.5) show an excess of interaction signatures in QSOs 2, that supports that at least some of the obscuration is due to the interaction, with the interstellar medium (ISM) also contributing to partly obscure the central regions, specially at higher redshifts (z$\sim$3--5) \citep{damato_dust_2020}. The influence of mergers was seen by \citet{storchi-bergmann_bipolar_2018}, who used a small sample of QSOs\,2 taken from the bigger sample of \citet{reyes_space_2008} and found an excess of interacting systems that showed evidence of outflows using HST imaging data. A follow-up to that work was presented by \citet{dallagnol_de_oliveira_gauging_2021}, where optical IFU data trom GMOS revealed the spatial distribution of outflowing gas in those same quasar hosts under interaction. In order to investigate if the interactions were indeed the origin of the nuclear activity in these QSO 2's, \citet{araujo_nuclear_2023} used a larger sample of $\approx400$ QSO\,2s, also drawn from \citet{reyes_space_2008} and matched them with control galaxies. Both were examined using SDSS and HST images and confirmed an excess of companions and asymmetric hosts for the QSOs\,2 due to mergers relative to control galaxies. A similar result was found by \citet{pierce_galaxy_2023}, which also found an excess of disturbed QSO\,2 hosts with  respect to non-AGN counterparts, concluding for a majority of pre-merging systems in the observed interactions.

Despite these results suggesting an evolutionary pathway between 
QSO 2's and 1's, \citet{shangguan_testing_2019} have found no evidence for this evolution in their previous work that investigated the gas content of QSO 1's as compared to QSO 2's: they found no difference between them an also no gas deficit with respect to non-active galaxies (under the assumption that the deficit would be due to feedback from the AGN). However, this work matched the type 1 and type 2 sources both in redshift and \oiii\ $\lambda$5007 luminosity. As this line is an indicator of the AGN luminosity, this selection may bias their result as they already selected the sources matching their gas content and AGN luminosity.

At higher redshifts, \citet{hamann_extremely_2017,perrotta_erqs_2019} have found a population of red quasars at $z > 2$ (from SDSS) that show type 1 emission lines in the UV, as well as a continuum that is flat/blue, while having red i-W3 colors. These sources show evidence of outflows due to blueshifted broad \oiii\ components.
This kind of object is consistent with a transitional stage between an obscured type 2 and a bluer type 1 QSO, because the obscuration found can be related to the intense outflows in the line-emitting regions.

At lower redshifts, a recent study by \citet{fawcett_striking_2023} investigates the relation between red quasars and radio emission using DESI optical spectra and LOFAR 144 MHz data. They have found a positive correlation between extinction and radio loudness, possibly due to shocks, suggesting the presence, in these sources, of a ``blowout'' phase in which the nucleus ejects the reddening material, revealing an intrinsic bluer source. In both cases, outflows of material from the nucleus seems related to the transition between a red nucleus (i.e. a type 2 QSO) into a blue one (i.e. a type 1 QSO).

Since recent results suggest an evolutionary scenario for the QSO population, our goal in this work is to investigate this further using optical spectroscopy data. Our aim is to compare and quantify the similarities and differences between type 1 and type 2 QSO populations with regard to their continua and emission line properties of the NLR. This will be done via two papers. In this first paper we present the sample, analysis methodology and results for the excitation of the NLR gas. In a second paper, we will present the results for the NLR gas kinematics.

Through this work, we adopted a Flat $\Lambda$CDM cosmology, using H$_0$\,=\,67.4\,\kms\,Mpc$^{-1}$, $\Omega_\Lambda$\,=\,0.685 and $\Omega_m$\,=\,0.315 \citep{planck_collaboration_planck_2020}.

\section{Sample and Data}

Since our goals are to compare types 1 and 2 QSO physical properties, the first step is to select the two samples, as explained below.

\begin{figure}
    \centering
    \includegraphics[width=\columnwidth]{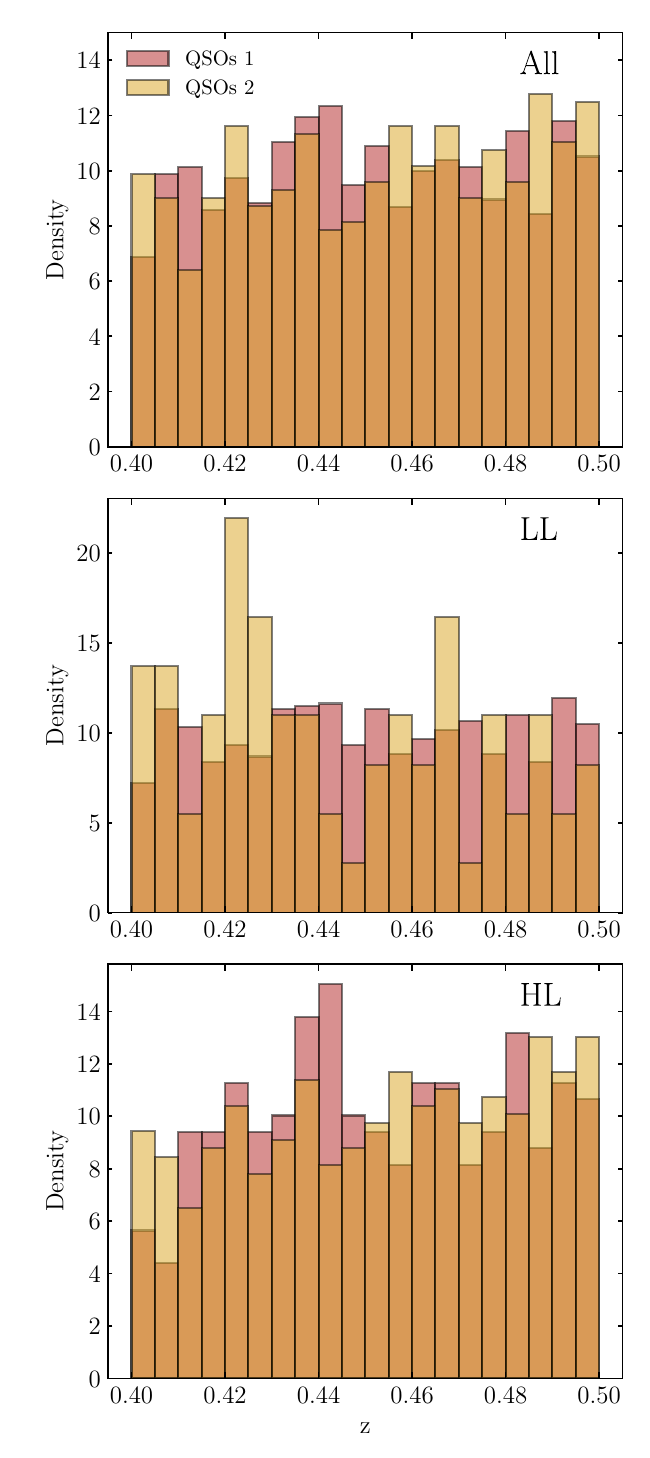}
    \caption{The normalized redshift $z$ distributions of the 3 QSO samples used in this study in the range $0.4 \le z \le 0.5$. In the upper panel, we show the full type 1 and 2 QSO samples as red and yellow histograms, respectively. In the middle panel, we show the QSO distribution for the LL sample, and in the lower panel, for the HL sample. We performed the two-sample KS test and found that the p-values are $\gtrsim$ 0.05 in all cases, meaning that the distributions are similar.}
    \label{fig:hist_z}
\end{figure}

\subsection{Type 2 QSOs}

\subsubsection{Reyes sample}\label{reyes_sample}
One of the most studied nearby QSO 2 samples is that of \citet[]{reyes_space_2008}, comprising 887 objects with redshifts of up to z=0.83, taken from the SDSS DR6. These sources were selected by those authors for presenting \oiii\ luminosity higher than the threshold L(\oiii)$\,>\,10^{8.3}$\,L$_\odot$. This sample has been used in a number of studies to investigate the relation between the the AGN and their host galaxies due to their proximity and high luminosity and the fact that, being of type 2, avoid the effect of the glare of the AGN continuum in the observations \citep[e.g.][and references therein]{fischer_minor_2015, fischer_hubble_2018, dallagnol_de_oliveira_gauging_2021, villar_martin_interactions_2021}. We selected a subsample of these objects in the redshift range $0.4 \leq z \leq 0.5$, as this interval corresponds to spectra covering the most important emission lines to characterize the NLR (e.g. \oiiid, \hb, \neiii$\lambda$3869, \nev$\lambda$3426). We have also selected for our analysis only the spectra presenting SNR $\gtrsim$ 10.

The resulting subsample used in this work -- hereafter Reyes sample, is comprised of 137 QSOs 2. Of these, 64 QSOs 2 present  L(\oiii)$\,>\,10^{9}$\,L$_\odot$, which is the lower limit of the \oiii\ luminosity of another QSO 2 sample used in this work -- the Yuan sample,  described below.

\subsubsection{Yuan sample}
Since the release of the Reyes sample, a newer and larger type 2 QSO sample was published by \citet[]{yuan_spectroscopic_2016}. This sample consists of 2758 type 2 QSOs from SDSS DR16 with redshifts of up to $z \approx 1$ selected mainly according to the rest equivalent width (REW) of the \oiii$\lambda5007$ emission line, which correlates with the line luminosity. Objects were selected for presenting REW(\oiii)\,$>100$\,\AA, which corresponds roughly to L(\oiii)$\,>\,10^{9}$\,L$_\odot$ \citep{yuan_spectroscopic_2016}. 

By selecting targets within the same redshift range of $0.4 \leq z \leq 0.5$, and also with corresponding SDSS spectra obeying the SNR $\gtrsim$ 10 criterium, the final sample is comprised of 551 unique objects -- hereafter the Yuan sample.

Our final type~2 sample is the combination of the above subsamples (Reyes + Yuan), resulting in 688 sources (hereafter QSO~2 sample).

\subsection{Type 1 QSOs}

The sample of type 1 QSOs was drawn from the catalog of \citet{lyke_sloan_2020}. It contains all spectroscopically confirmed type 1 QSOs from SDSS up to its DR16 -- the Data Release 16 Quasar catalog (DR16Q). DR16Q contains 750414 objects with spectroscopic redshifts $z$ in the range $0.0007 < z < 7.02$. The data comprise the rest-frame ultraviolet (UV) to optical spectral properties of these Quasars. By applying the same selection criteria as in \S~\ref{reyes_sample} we were left with 2827 sources. 

Since DR16Q may contain not only types 1 QSOs but also some type 2, and is drawn from a larger sample (more recent DR), it is necessary to remove possible duplicates, in particular type 2 sources. Therefore, we followed a two-fold step:

\begin{enumerate}
    \item[{\bf i) Cross-match the catalogues:}] This search identified 71 common objects between the Reyes and the DR16Q catalog, as well as 2 common objects between the Yuan and DR16Q, which revealed the existence of type 2 QSOs within the \citet{lyke_sloan_2020} sample. These objects were flagged and removed from the DR16Q subsample. We also found duplicates between the type 2 catalogs: 44 objects were found in both Reyes and Yuan catalogs, therefore resulting in them being flagged and removed from the Yuan subsample. This step resulted in each duplicate object being analyzed only once in this work.

    \item[{\bf ii) Visual inspection for additional QSO 2s:}] After removing the objects in common for both QSO 2 samples, we have visually inspected the remaining objects in the DR16Q sample to find additional possible QSOs 2, falgging them as such according with the following criteria: (i) a flat spectral continuum, with no power-law component; (ii) absence of broad components in the \mgii\ and \hb\ lines;
    (iii) absence of \feii\ multiplet components. From this, we recovered 311 additional type 2 QSOs and removed them from the QSO 1 sample, excluding them from this study. Our final QSO 1 sample has 1541 sources (hereafter QSO~1 sample).
    
\end{enumerate}

\subsection{Luminosity range} 

After fitting the emission line profiles (see \S~\ref{sec:specfit}), we have compared the derived properties of QSOs~1 with those of QSOs~2 in different luminosity regimes. This was motivated by the fact that the two QSO~2 sub-samples have a distinct lower luminosity limit (L$_{\textrm{\oiii}}$\,>\,$10^{8.3}$\,L$_\odot$ and L$_{\textrm{\oiii}}$\,>\,$10^{9}$\,L$_\odot$ for the Reyes and Yuan samples, respectively). We have thus separated our sample in two luminosity subsamples as follows: 

\begin{enumerate}
    \item A {\bf ``low luminosity''} sample (LL), comprised of 1222 QSOs 1 and 73 QSOs 2 originating from the Reyes sample, with \oiii\ luminosities within the range $10^{8.3}$\,L$_\odot$\,<\,L$_{\textrm{\oiii}}<10^{9}$\,L$_\odot$;
   
    \item A {\bf ``high luminosity''} sample (HL), with 319 type 1 QSOs to be compared with 615 type 2, from the combined Reyes (64 objects) and Yuan (551 objects) samples, with L$_{\textrm{\oiii}} > 10^{9}$ L$_\odot$.
\end{enumerate}

The redshift distribution of our final QSO~1 and QSO~2 samples is shown on the top panel of Figure\,\ref{fig:hist_z}, while the low and high luminosity samples are presented in the middle and lower panels, respectively.

Due to sample size differences, we have also performed a KS-test on the histograms of Figure\,\ref{fig:hist_z}, and found p-values larger than 0.05, revealing the similarity between both QSO 1 and 2 populations in all luminosity ranges.

\section{Methodology}

\subsection{Spectral fits}\label{sec:specfit}

\begin{figure*}
    \centering
    \includegraphics[width=0.96\textwidth]{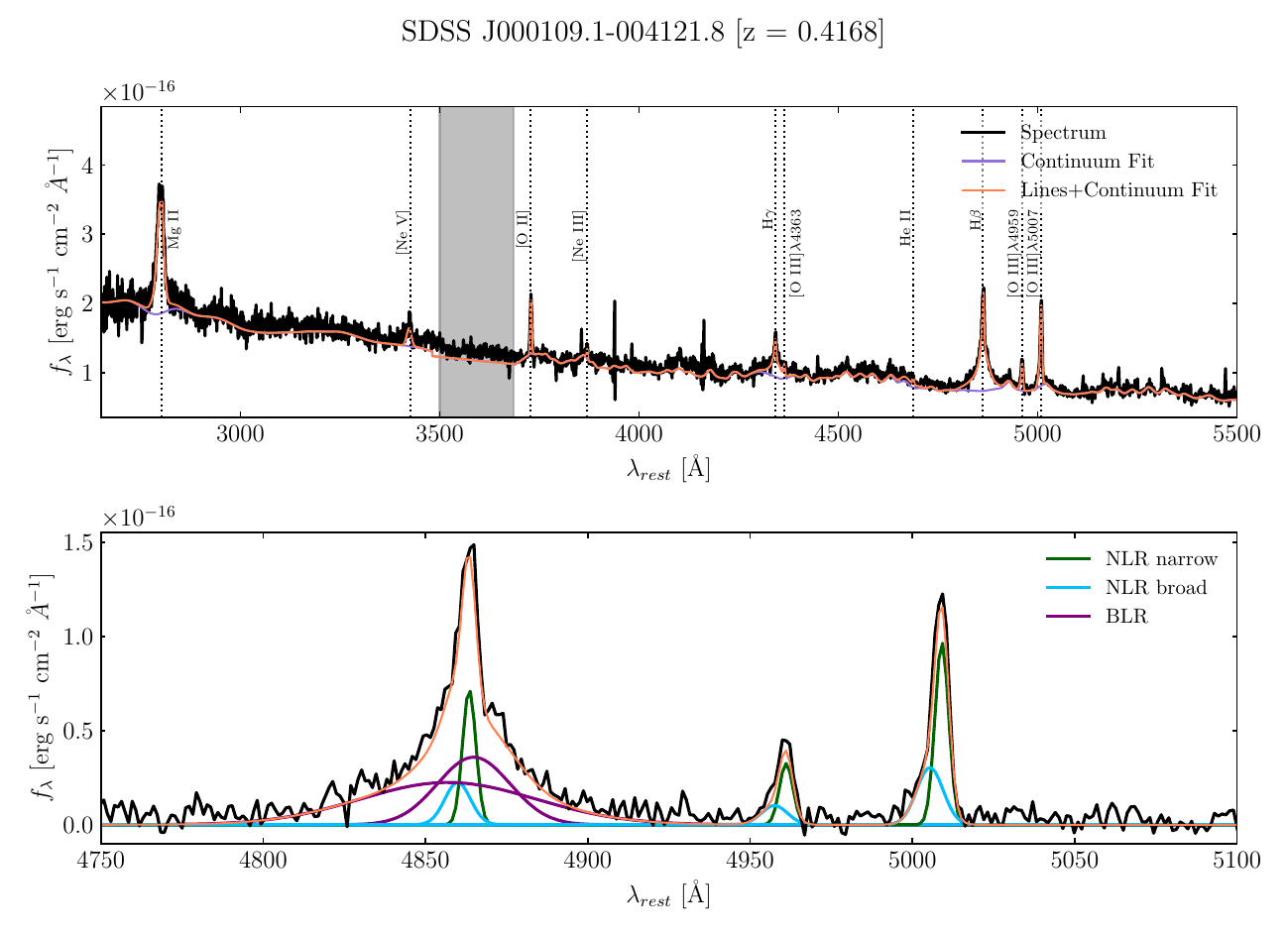}
    \caption{Example of a type 1 QSO spectrum. \textbf{Top}: the redshift-corrected flux density in black; in light blue the PyQSOFit fitted continuum and in orange, the continuum and emission lines fit. As vertical black dashed lines, we show the rest wavelengths of the \mgii, \nev, \oii, \neiii, \hg, \oiii$\lambda4363$, \hei, \hb\ and \oiiid\ emission lines, from left to right. In shaded gray, we denote the spectral window without \feii\ template data. \textbf{Bottom}: zoomed-in emission line fitting around the \hb+\oiiid\ complex.  In black, we denote the continuum-subtracted flux density. The NLR$_\mathrm{n}$ component is shown in green, the NLR$_\mathrm{b}$ in blue and the BLR components in magenta. In orange, we show the full fitted profiles.}
    \label{fig:spec_fit_qso1}
\end{figure*}

\begin{figure*}
    \centering
    \includegraphics[width=0.96\textwidth]{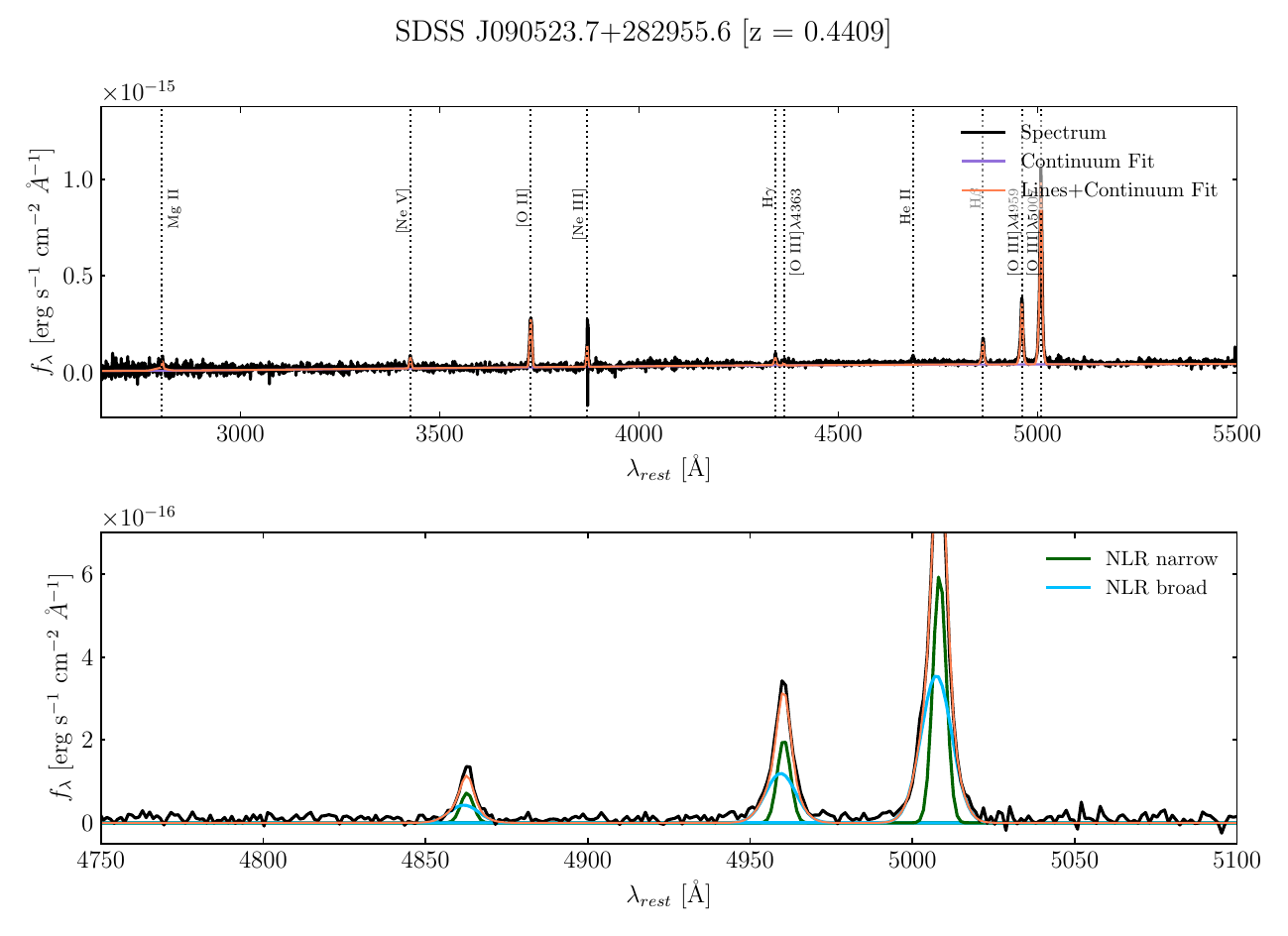}
    \caption{Example of a type 2 QSO spectrum. \textbf{Top}: the redshift-corrected flux density in black; in orange, the continuum and emission lines fit. As vertical black dashed lines, we show the rest wavelengths of the \mgii, \nev, \oii, \neiii, \hg, \oiii$\lambda4363$, \hei, \hb\ and \oiiid\ emission lines, from left to right. \textbf{Bottom}: zoomed-in emission line fitting around the \hb+\oiiid\ complex. In black, we denote the continuum-subtracted flux density. The NLR$_\mathrm{n}$ component is shown in green and the NLR$_\mathrm{b}$ in blue. In orange, we show the full fit.}
    \label{fig:spec_fit_qso2}
\end{figure*}

The Quasar optical and NUV spectra of our sample is a combination of strong emission lines -- produced by ions such as Mg$^{+}$ and O$^{2+}$ -- blended with other fainter lines including weak emission lines from the \feii\ multiplet and a continuum component produced by thermal emission from the SMBH accretion disk as well as free-bound electronic transitions from Hydrogen. In order to study the physical origin of the continuum and characterize the profiles of the emission lines, considering the presence of these complex continua components, we used the PyQSOFit library \citep{guo_pyqsofit_2018}. This is a well-known code previously used, for example, in the derivation of the Quasar properties listed in the \citet{wu_catalog_2022} catalog, which fits Gaussian components to the line profiles together with the continuum. For the derivation of the parameter uncertainties, it also performs Monte Carlo resampling, for which we opted to do 200 iterations for each spectrum.

\subsubsection{Continuum and \feii\ multiplet}

The continuum is fitted using three components: a power-law f$_\lambda$ \,$\propto \lambda^{b}$, a 3$^\mathrm{rd}$ degree polynomial and a Balmer continuum component, to account for the electron recombination emission. In addition to the continuum, PyQSOFit also fits the \feii\ multiplet component, doing this separately for the ``UV\,\feii'' region and the ``optical\,\feii'' region, in order to better fit the blended iron lines.

\subsubsection{Emission Lines}

\begin{table*}
\caption{Initial parameters given to PyQSOFit for fitting the emission lines. From left to right, the columns list: (1) the identification of the emission line component; (2) the rest wavelength ($\lambda$); (3) the wavelength window used for each fitting group ($\lambda_\mathrm{window}$) -- each group separated by the horizontal lines; (4) the maximum velocity offsets for each Gaussian component fitted ($v_\mathrm{off}$); (5) the initial velocity dispersion given to each component ($\sigma_\mathrm{ini}$); (6) and (7)  the velocity dispersion limits for each component ($\sigma_\mathrm{inf}$, $\sigma_\mathrm{sup}$). Marked with a $\dagger$, are the BLR components considered to be absent in type 2 QSOs.}
\label{tab:pyqsofit_params}
\begin{tabular}{l|cccccc}
\hline \hline
\multicolumn{1}{c|}{Component} & $\lambda$ [\AA] & $\lambda_\mathrm{window}$ [\AA] & $v_\mathrm{off}$ [\kms] & $\sigma_\mathrm{ini}$ [\kms] & $\sigma_\mathrm{inf}$ [\kms] & $\sigma_\mathrm{sup}$ [\kms] \\ \hline
\oiii$\lambda5007$ NLR$_\textrm{n}$ & 5008.24 & \multirow{11}{*}{4250 -- 5100} & 600 & 250 & 100 & 600 \\
\oiii$\lambda5007$ NLR$_\textrm{b}$ & 5008.24 &  & 1000 & 900 & 200 & 1200  \\
\oiii$\lambda4959$ NLR$_\textrm{n}$ & 4960.29 &  & 600 & 250 & 100 & 600  \\
\oiii$\lambda4959$ NLR$_\textrm{b}$ & 4960.29 &  & 1000 & 900 & 200 & 1200  \\
\hb\ NLR$_\textrm{n}$ & 4862.68 &  & 600 & 300 & 100 & 600  \\
\hb\ NLR$_\textrm{b}$ & 4862.68 &  & 1000 & 900 & 200 & 1200  \\
\hb\ BLR$^\dagger$ & 4862.68 &  & 3000 & 1000 & 900 & 3900  \\
\oiii$\lambda4363$ NLR$_\textrm{n}$ & 4363.21 &  & 600 & 250 & 100 & 600  \\
\hg\ NLR$_\textrm{n}$ & 4341.69 &  & 600 & 250 & 100 & 600  \\
\hg\ NLR$_\textrm{b}$ & 4341.69 &  & 1000 & 900 & 200 & 1200  \\
\hg\ BLR$^\dagger$ & 4341.69 &  & 3000 & 1000 & 900 & 3900  \\ \hline
\neiii$\lambda3869$ NLR$_\textrm{n}$ & 3868.76 & \multirow{2}{*}{3690 -- 3900} & 480 & 300 & 0 & 500 \\
\oii$\lambda3728$ NLR$_\textrm{n}$ & 3728.38 &  & 480 & 300 & 0 & 500 \\ \hline
\nev$\lambda3426$ NLR$_\textrm{n}$ & 3426.85 & 3380 -- 3460 & 480 & 300 & 0 & 500  \\ \hline
\mgii$\lambda2800$ NLR$_\textrm{n}$ & 2803.53 & \multirow{3}{*}{2700 -- 2900} & 600 & 250 & 100 & 1000 \\
\mgii$\lambda2800$ NLR$_\textrm{b}$ & 2803.53 &  & 3000 & 900 & 200 & 2000 \\
\mgii$\lambda2800$ BLR$^\dagger$ & 2803.53 &  & 3000 & 1000 & 900 & 6000\\
\hline
\end{tabular}
\end{table*}

SDSS spectra of QSOs in the $0.4 \leq z \leq 0.5$ interval cover a spectral window containing two particularly conspicuous  sets  of lines -- \mgii$\lambda2800$ and \oiiid\ emission lines -- as well as other emission lines between them, such as \nev$\lambda3427$, \oii$\lambda3728$, \neiii$\lambda3869$ and the Balmer lines such as \hg\ and \hb. To account for the emission originated in the NLR and in the BLR, we used a combination of Gaussian profiles in the fit: 2 for the NLR component -- one narrow and one broader -- and 2 broad components for the BLR. Since forbidden emission lines are not present in the BLR -- as their critical densities are lower than the typical BLR gas densities, we excluded the BLR component for the \nev, \oii, \neiii\ and \oiiid\ lines. We also excluded the broad NLR component for the \nev, \oii\ and \neiii\ lines due to their relatively low strengths relative to the noise level, and the fact that the former two have wavelengths close to those of the edges of the \feii\ UV and optical templates precluding their detection.

Fitting the emission lines with PyQSOFit requires, prior to anything, the definition of \textit{fitting groups} of lines, such that the emission lines within the spectral window of a group are fitted simultaneously and their profile parameters (such as relative amplitude -- in the case of doublets -- velocity shift and velocity dispersion) may be tied together. In this work, we defined 4 different fitting groups due to spectral proximity and physical coupling of the emission line properties: (1) the \mgii\ profile; (2) the \nev\ line; (3) the \oii\ and \neiii\ lines  -- although presenting different ionization potentials, these lines are frequently weak, and we can recover them better coupling their NLR profiles; (4) the  \hg\ and \oiii$\lambda4363$ lines and the complex composed by the \hb\ line and \oiii\ doublet. In Table\,\ref{tab:pyqsofit_params}, we list the spectral windows adopted for each fitting group as $\lambda_\mathrm{window}$, and separate the different groups with horizontal lines. After defining the fitting groups, we input the other parameters for the fitting: the rest wavelength of the lines\footnote{Emission line data taken from: \url{https://www.nist.gov/pml/atomic-spectra-database}} ($\lambda$); the maximum velocity offsets for each component ($v_\mathrm{off}$), the initial velocity dispersion ($\sigma_\mathrm{ini}$) and the velocity dispersion lower and upper bounds ($\sigma_\mathrm{inf}$, $\sigma_\mathrm{sup}$). The values assigned for the above initial parameters for each emission line are listed in Table\,\ref{tab:pyqsofit_params}. In order to achieve physical results, we constrained the velocity centroid and velocity dispersion for the following components to be the same, as they should originate in the same physical region:

\begin{enumerate}
    \item the \oiii\ doublet narrow components and the Balmer lines narrow components;
    \item the \oiii\ doublet NLR broad components and the Balmer lines NLR broad components;
    \item the Balmer lines BLR individual components;
\end{enumerate}
 
\noindent therefore giving the same initial parameters for each tied component, similarly to what has been done in previous works \citep[e.g.][]{ruschel-dutra_agnifs_2021, dallagnol_de_oliveira_gauging_2021, wu_catalog_2022, hauschild-roier_gas_2022, riffel_mapping_2023}. Since the \oiii\ lines are doublets, we also constrained PyQSOFit to scale the 5007\,\AA\ components amplitudes as 2.98 times of those of the 4959\,\AA\ line \citep{osterbrock_astrophysics_2006}.

Since QSOs 2 do not possess BLR components in their spectra, we did not include the BLR components (marked with a $\dagger$ in Table\,\ref{tab:pyqsofit_params}) in their fitting procedure. For the components present in both type 1 and 2 QSOs, we used the same constraints.

Due to numerical degeneracy of the emission line fitting procedure, we opted to use the relative amplitude between the narrow and broad components of the \oiiid\ doublet as a proxy for the NLR profiles of the Balmer lines. In order to do this, we first executed PyQSOFit with the initial parameters from Table\,\ref{tab:pyqsofit_params} to obtain the individual components for the \oiiid\ doublet. Next, we fixed the \oiii\ emission line parameters to match those obtained in the previous step and constrained the Balmer lines' narrow and broad NLR components amplitudes to match those of the \oiii. With this additional constraint between the Balmer and \oiiid\ lines, we finally fitted the spectra.

We show as an example of this procedure a type 1 and a type 2 QSO spectrum (in black), overlapped with the continuum (in light blue) and emission line (in orange) fitting from PyQSOFit in the upper panels of Figures\,\ref{fig:spec_fit_qso1} and \ref{fig:spec_fit_qso2}, respectively, within 2600\,--\,5500\AA\ in rest frame. We also mark the rest wavelengths of the fitted emission lines (\mgii, \nev, \oii, \neiii, \hg, \oiii$\lambda4363$, \hei, \hb\ and \oiiid) as dotted black vertical lines. In the lower panels of Figures\,\ref{fig:spec_fit_qso1} and \ref{fig:spec_fit_qso2}, we also show a zoomed-in view of the \hb+\oiiid\ complex region. In these Figures, the continuum-subtracted flux density is shown in black, while each individual Gaussian component is in a different color: green for the NLR$_\mathrm{n}$, red for the NLR$_\mathrm{b}$ and blue for the two BLR components. The full fitted profile is shown in orange. For the type 1 QSO spectrum in the top of Figure\,\ref{fig:spec_fit_qso1}, we also shade in grey the spectral window where there is no data  for the \feii\ template to be used in the PyQSOFit fitting.

In the analysis of the data, we considered as significant only emission lines with luminosities higher than 3 times the ``noise luminosity'', defined as the area of a Gaussian profile with amplitude equal to the spectral noise given by the local continuum RMS, and FWHM corresponding to the mean spectral resolution $\Delta\lambda = 1.25$\,\AA. This procedure was adopted in order to avoid considering as emission lines possible spurious features originated from fluctuations due to the noise, and adds an additional constraint of SNR > 3 per emission line component.

\subsection{Electronic temperature}

The \oiii$\lambda$4363 line forms a triplet together with the \oiiid\ doublet, since the first transition shares its lower level with the upper level of the other two. This physical relation allows the determination of the electronic temperature (T$_e$) of the ISM where these lines are produced. We used the software PyNeb \citep{luridiana_pyneb_2015} to determine T$_e$ using the $\lambda$4363/$\lambda$5007 line ratio, assuming an electronic density n$_e$\,=\,100\,cm$^{-3}$ throughout this work, as this diagnostic is insensitive to densities up to n$_e \sim 10^4$\,cm$^{-3}$ \citep{osterbrock_astrophysics_2006}. Due to high uncertainties on measuring the \oiii$\lambda$4363 line for the LL sample, we calculated T$_e$ only for the HL sample. These results are presented and further discussed in Section\,\ref{sec:te_res_highL}.

\section{Results and Analysis - Low Luminosity} \label{sec:res_lowL}

\begin{table}
    \centering
    \caption{Median values for each QSO type for the line luminosities, line ratios and electronic temperature parameters (discussed in Section\,\ref{sec:res_lowL}) for the LL sample. The median values are shown in the following histograms as dashed lines -- for QSOs 1 -- and dotted lines -- for QSOs 2. Units are as follow: luminosities are in \lsun and masses are in \msun. We also include the KS test p-value for each parameter.}
\begin{tabular}{lccc}
\hline \hline
Parameter & QSO 1 & QSO 2 & p$_\mathrm{KS}$ \\
\hline
log(L\oiii$\lambda$5007),n & 8.34 & 8.33 & 5.72$\times 10^{-2}$ \\
log(L\oiii$\lambda$5007),b & 8.28 & 8.44 & 4.75$\times 10^{-4}$ \\
log(L\oiii$\lambda$5007) & 8.63 & 8.70 & 6.72$\times 10^{-3}$ \\
log(L \hb),n & 8.10 & 7.94 & 2.81$\times 10^{-5}$ \\
log(L \hb),b & 8.12 & 7.84 & 7.23$\times 10^{-11}$ \\
log(L \hb) & 8.10 & 7.79 & 2.81$\times 10^{-5}$ \\
\oiii$\lambda$5007 n/b & 1.12 & 0.76 & 3.50$\times 10^{-5}$ \\
\hb\,NLR n/b & 1.02 & 1.22 & 4.78$\times 10^{-2}$ \\
log(L\oii) & 8.05 & 8.05 & 4.26$\times 10^{-2}$ \\
log(L\neiii) & 7.91 & 7.74 & 7.84$\times 10^{-10}$ \\
log(L\oiii$\lambda$4363) & 7.96 & 7.39 & 3.90$\times 10^{-12}$ \\
log(L\nev) & 8.09 & 7.50 & 2.24$\times 10^{-39}$ \\
\nev/\neiii & 1.48 & 0.67 & 9.18$\times 10^{-13}$ \\
\neiii/\oii & 0.60 & 0.38 & 1.14$\times 10^{-7}$ \\
\oiii$\lambda$5007/\oii & 4.25 & 4.34 & 7.55$\times 10^{-2}$ \\
\oiii$\lambda$5007/\hb & 3.89 & 10.31 & 2.81$\times 10^{-5}$ \\
\oiii$\lambda$5007/\hb,n & 1.77 & 7.90 & 1.59$\times 10^{-10}$ \\
\oiii$\lambda$5007/\hb,b & 2.41 & 6.72 & 3.33$\times 10^{-14}$ \\
log(M \hb) & 7.66 & 7.35 & 2.81$\times 10^{-5}$ \\
\hline
\end{tabular}
\label{tab:median_exc_lowL}
\end{table}

\subsection{Emission line luminosities}

\begin{figure*}
    \centering
    \includegraphics[width=0.85\textwidth]{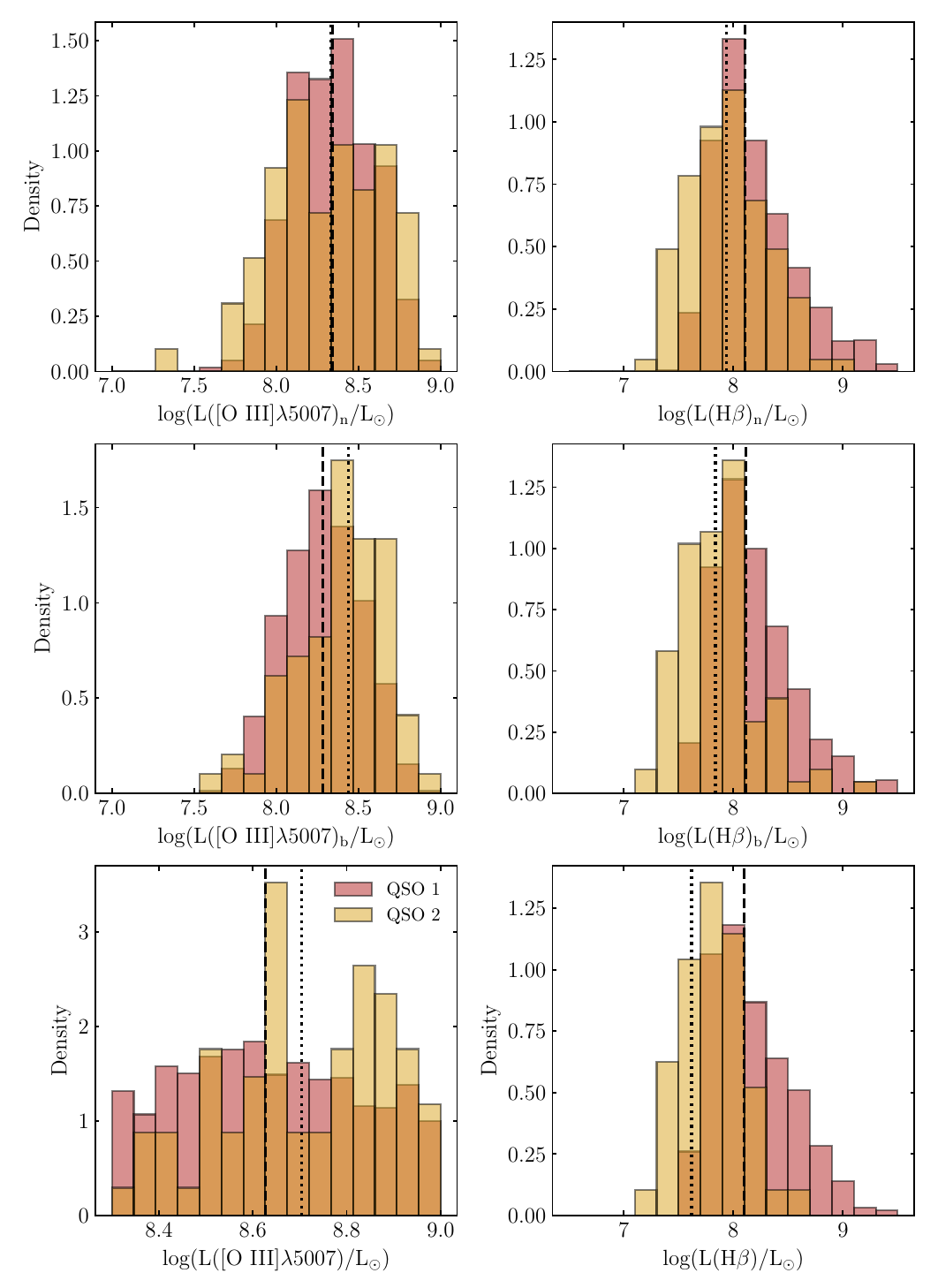}
    \caption{Normalized distributions for the narrow line luminosities NLR$_\textrm{n}$ (top), broad NLR$_\textrm{b}$ (middle) and total NLR luminosities (bottom), for the \oiii$\lambda$5007 (left) and \hb\ emission lines (right) for the LL sample. The QSO 1 and QSO 2 populations are shown as red and yellow histograms, respectively, showing the median of each population as black dashed (QSOs 1) or dotted (QSOs 2) vertical black lines.}
    \label{fig:hist_L_comp_lowL}
\end{figure*}

In this section, we present and discuss the results regarding the distributions of the gas luminosity in the different emission lines, comparing the results for QSOs 1 and 2 for the low-luminosity sample. For the emission lines with two components, we also compare separately results for the narrow component, identified with a subscript {\it n}, e.g. NLR$_n$ and for the broad component NLR$_b$. The median values of the distributions as well as the corresponding KS-test p-values (p$_\mathrm{KS}$) are shown in Table\,\ref{tab:median_exc_lowL}. 

For robustness, and to verify if the different numbers of sources in the QSO 1 and 2 samples does not affect the results, we performed a random resampling test of the data, with 1000 iterations, matching the larger QSO 1 sample size to the smaller QSO 2 sample size, computed their median values and performed KS-tests. In these tests, we found that the medians deviate $\lesssim\,1\%$ from values found for the full samples, and that the KS-test results remain the same.

\medskip
\noindent{\bf \oiii\ and \hb\ --} The panels of Figure\,\ref{fig:hist_L_comp_lowL} show the observed luminosity distributions in each NLR component and for the whole profile, for the \oiii$\lambda$5007 (left) and \hb\ lines (right). The top column panels refer to the narrow component, the middle one to the broad ones, and the bottom panels show the distributions for the total profiles. 

Fig. \,\ref{fig:hist_L_comp_lowL} and Table\,\ref{tab:median_exc_lowL} show that, although the distribution of the \oiii\ luminosities for the NLR narrow component for QSOs 2 and 1 are similar (as revealed by their similar medians and a KS-test p-value\,$>0.05$), the broad component shows a bimodality, with the QSOs 2 distribution having a higher fraction of higher luminosity objects than the type 1 QSOs, as traced by their respective medians, that indicate a $\approx$\,45\% higher value for the type 2. Regarding the sum of the broad and narrow components, shown in the bottom panels, as a consequence of the results for the broad component, the fraction of objects with the highest luminosities is higher for the type 2 QSOs. 

We note that, if we consider all the QSOs 2 from the Reyes sample (without dividing into LL and HL samples) and compare their L\oiii\ distribution with that of the QSOs 1, the overall L\oiii\ distributions are distinct with median values of 8.95 (type 2) and 8.71 (type 1), implying a 75\% higher L\oiii\ for the type 2's. This remark is important because the division between LL and HL QSOs lies approximately "on top" of the peak and median value of the QSOs 2 distribution, as probed by the Reyes sample. The higher peak (and mean) L\oiii\ luminosity for QSOs 2 than for QSOs 1 indicates distinct AGN properties, supporting the evolutionary scenario, and is in line with previous results reported for QSOs at "cosmic noon" \citep{tozzi_super_2024}.

For the \hb\ components, the difference in the distributions of QSOs 1 and 2 is opposite to that of \oiii, with a higher fraction of higher luminosity sources in QSOs 1 than in QSOs 2, and this is observed both in the narrow and broad NLR components and the full NLR profile, with the median of the total luminosities of the two \hb\ components of the QSOs 1 being twice the value of that for the QSOs 2. 

\begin{figure*}
    \centering
    \includegraphics[width=0.85\textwidth]{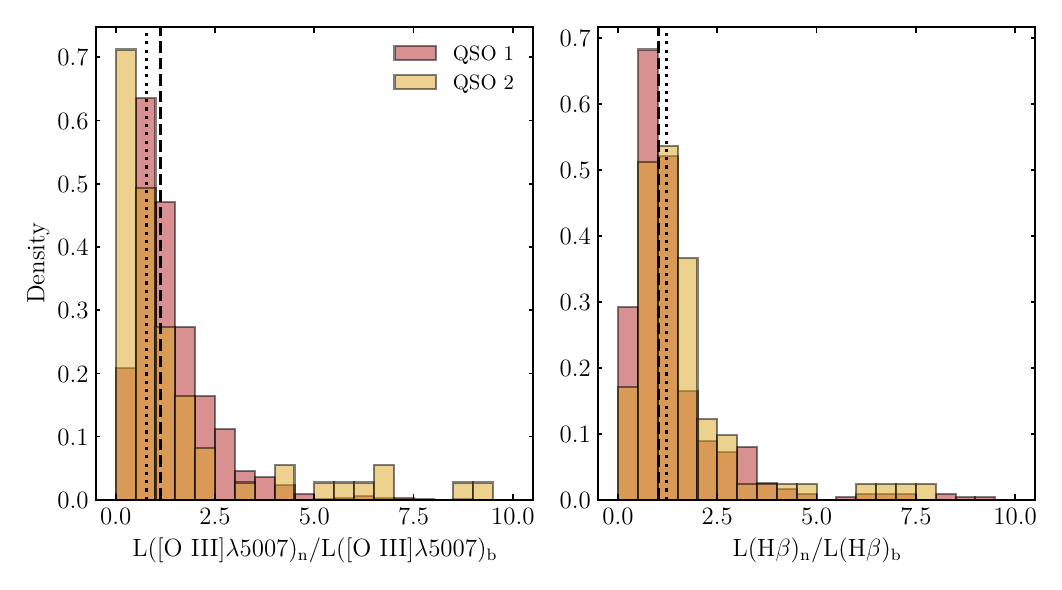}
    \caption{Normalized \oiii$\lambda$5007 NLR$_\mathrm{n}$/NLR$_\mathrm{b}$ (left) and \hb\ NLR$_\mathrm{n}$/NLR$_\mathrm{b}$ (right) components ratios, for the LL sample. The QSO 1 and QSO 2 populations are shown as red and yellow histograms, respectively, showing the median of each population as black dashed or dotted vertical black lines.}
    \label{fig:hist_na_br_lowL}
\end{figure*}

Beyond investigating separately the luminosities of the NLR$_\mathrm{n}$ and NLR$_\mathrm{b}$ components, we also investigate the luminosity ratio between the narrow and broad components. We show the results for the \oiii$\lambda$5007 and the \hb\ lines in Figure\,\ref{fig:hist_na_br_lowL}. For \oiii\, the ratios $narrow/broad$ for the QSOs 2 tend to be slightly lower than those for the QSOs 1, with the median ratio for the QSOs 2 being 0.76 and for the QSOs 1 1.12, indicating a somewhat stronger contribution of the broad NLR component in type 2 relative to type 1 sources.

The higher luminosity in the $broad$ NLR component of \oiii\ in QSOs 2 indicates a difference in the NLRs of QSOs 1 and QSOs 2 that cannot be attributed to the Unified Model. Since the broad NLR component is usually identified with kinematically disturbed regions -- e.g. outflows, a possible interpretation is the higher incidence of these phenomena in type 2 relative to type 1 sources, such as presented by \cite{tozzi_super_2024}, in which a broader \oiii\ profile in type 2 QSOs (in respect to type 1s) is found to trace radiation pressure-driven outflows more efficiently. Besides outflows due to the AGN, another possibility is a higher incidence of interactions in the host galaxies. This is supported by the fact that QSOs 2 -- and in particular those from the Reyes sample, show a high incidence of interacting companions \citep{storchi-bergmann_bipolar_2018, araujo_nuclear_2023, pierce_galaxy_2023}, both fueling the host galaxy with gas and contributing to the disturbed kinematics. This gas, that frequently extends to beyond the borders of the stellar component of the galaxy, emits strongly in \oiii, as shown in \citet{storchi-bergmann_bipolar_2018}.

Regarding H$\beta$, there is practically no difference in the ratios NLR$_\mathrm{n}$/NLR$_\mathrm{b}$ between the QSO 1 and 2 samples.

\begin{figure*}
    \centering
    \includegraphics[width=0.85\textwidth]{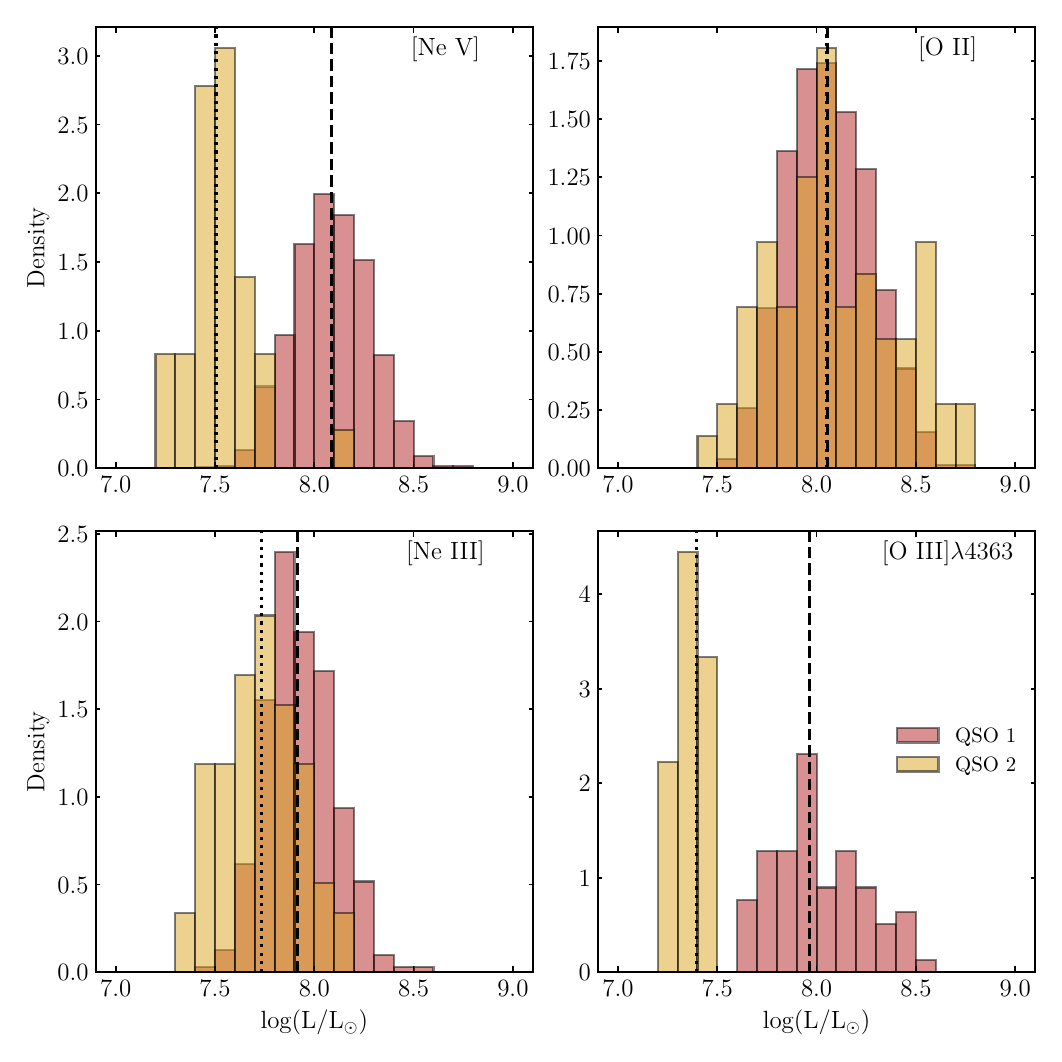}
    \caption{Normalized luminosity distributions of the \nev, \oii, \neiii\ and \oiii$\lambda$4363 emission lines for the QSO 1 and QSO 2 populations for the LL sample, shown as red and yellow histograms, respectively, showing the median of each population as black dashed or dotted vertical black lines.}
    \label{fig:hist_lum_lowL}
\end{figure*}

\medskip
\noindent{\bf \nev, \oii, \neiii\ and \oiii$\lambda$4363 --} The luminosity distributions for these lines are shown in the four panels of Figure\,\ref{fig:hist_lum_lowL}, where we present the medians as vertical dashed or dotted lines, for the QSOs 1 and QSOs 2, respectively. The numerical values of each median are presented in Table\,\ref{tab:median_exc_lowL}. \nev\ can be considered a coronal line for its high ionization potential of 97.11\,eV \citep[e.g.][]{rodriguez-ardila_complex_2017, rodriguez-ardila_700_2020}, thus a clear AGN signature. We find a strong bimodality in its luminosity distribution, with the type 1 QSOs being $\approx4$ times more luminous than in type 2 objects. For the less ionised Neon line, \neiii, a weaker bimodality is found, but with QSOs 1 still presenting higher luminosities. For \oii\ there is no difference between type 1 and 2's, but for \oiii$\lambda$4363, we observe again a bimodality, with type 1's being $\approx4$ times more luminous than type 2's.

The bimodalities of these higher excitation lines indicate we are seeing more regions of higher radiative excitation in type 1s with respect to the type 2 QSOs. A higher excitation in QSOs 1 could be consistent with the Unified Model of AGN \citep{antonucci_unified_1993}, in which we are seeing more of the central region of the QSOs 1 which is hidden in QSO 2s, thus being able to observe gas closer to the accretion disk, hotter and with higher ionization potentials \citep{osterbrock_astrophysics_2006}. Alternatively, in an evolutionary scenario, type 2's are in an earlier stage, more enshrouded in gas and dust that are later blown away by the AGN, clearing the view of the hotter and higher ionization regions in the type 1 phase.

\subsection{Emission-line ratios}
\label{sec:exc_res}

The analysis of the emission-line ratios can reveal the nature of the excitation source of these objects due to the different energies necessary to excite and ionize the atoms, in order to produce the observed ions. For the studied emission lines, the ionization potentials are as follows: for the O$^{+}$ and O$^{2+}$ ions they are 13.61 and 35.12\,eV, respectively; for Ne$^{2+}$ and Ne$^{4+}$ they are 40.96 and 97.11\,eV, respectively; and finally, the ionization energy for H$^{+}$ is 13.59\,eV. Therefore, the \oii\ and Balmer lines require similar ionization energies, while the \neiii\ and \oiii\ also have similar ionization potentials and \nev\ has an ionization potential almost 2.75 times higher than that of the \oiii\ lines. 

\begin{figure*}
    \centering
    \includegraphics[width=0.85\textwidth]{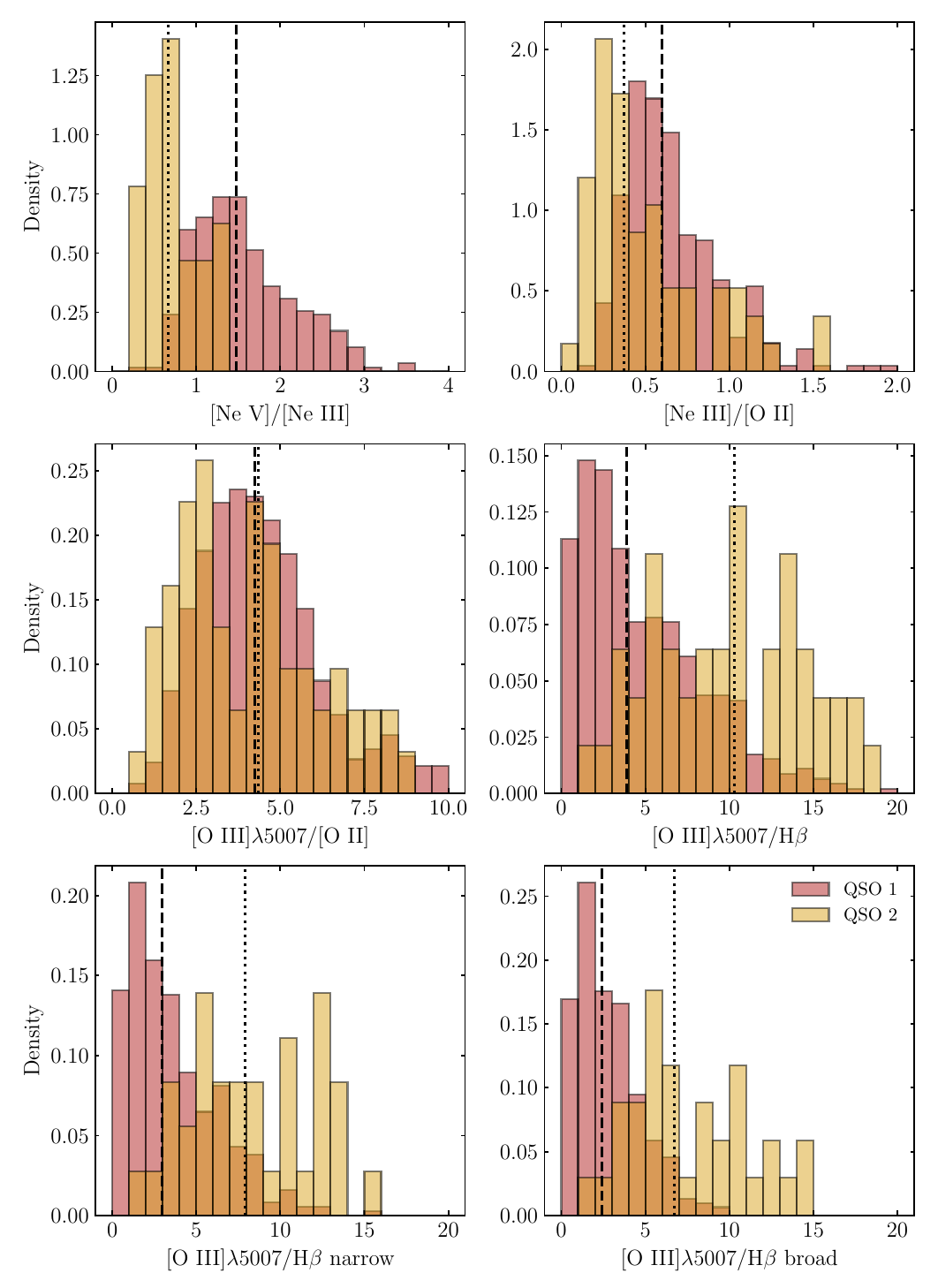}
    \caption{Normalized distributions of the \nev/\neiii, \neiii/\oii, \oiii$\lambda$5007/\oii\ and \oiii$\lambda$5007/\hb\ line ratios, also including the individual NLR components ratios for the latter, narrow (left) and broad (right). The distributions are for the QSO 1 and QSO 2 populations for the LL sample, shown as red and yellow histograms, respectively. The median of each population are shown as black dashed or dotted vertical black lines.}
    \label{fig:hist_ratios_lowL}
\end{figure*}

We present 4 different line ratio histograms in the panels of Figure\,\ref{fig:hist_ratios_lowL}: \nev/\neiii, \neiii/\oii, \oiii$\lambda$5007/\oii\ and the total and component-wise \oiii$\lambda$5007/\hb. 

\medskip 
\noindent{\bf \nev/\neiii\ -- } This ratio shows higher values, in general, in type 1 objects than in type 2, reaching up to $\sim3$ and $\sim1.5$, respectively. Due to the very high ionization potential of \nev, it should be produced closer to the AGN than \neiii, from a region better observed in QSOs 1 than in QSOs 2, according to the Unified Model - that posits that we see the AGN more pole-on in the type 1 QSOs. This is supported by the smaller difference between the \neiii\ luminosities for QSOs 1 and QSOs 2 and higher difference, with type 2's being much less luminous in \nev\ (see Figure\,\ref{fig:hist_lum_lowL}). 

\medskip
\noindent{\bf \neiii/\oii\ -- } This ratio reaches higher values for QSOs 1 than for QSOs 2, consistent again with more regions of higher excitation being seen in QSOs 1 than in QSOs 2.

\medskip
\noindent{\bf \oiii$\lambda$5007/\oii\ -- } This ratio is similar for the two QSO types, in spite of the fact that \oiii\ present a higher fraction of type 2s towards the high luminosity end.

\medskip 
\noindent{\bf \oiii$\lambda$5007/\hb\ -- } This ratio tends to be larger for QSOs 2 with respect to QSOs 1. This trend is also seen for the individual NLR components.

In the BPT diagrams \citep{baldwin_classification_1981} the \oiii/\hb\ ratio is used to discriminate Seyfert galaxies from LINERs: while Seyfert galaxies have typically \oiii$\lambda$5007/\hb$\geq10$, LINERs have \oiii$\lambda$5007/\hb$\leq3$ -- 5. From the distribution, there are few LINER-like line ratios among QSOs 2 (7 objects), but many among QSOs 1 (199 objects). The average ratio for QSOs 1 is 4, while it is about 10 for QSOs 2.

LINER nuclei can be due both to low-luminosity AGN and to regions being heated by shocks. As all our sources have high luminosities (we limited L(\oiii) to be higher than $10^{8.3}$\,\lsun), the most probable explanation for the
lower \oiii$\lambda$5007/\hb\ ratios in QSOs 1 is the presence of shocks \citep{dopita_spectral_1995}. This difference may indicate that QSOs 1 are intrinsically different from QSOs 2 or could be due to the fact that these shocks occur very close to the nucleus and are blocked from view in QSOs 2, in agreement with the Unified Model.

\subsection{Ionised gas mass}

We can use our measurements to obtain the ionised gas mass, which can be derived from the \hb\ luminosity via the following relation \citep{osterbrock_astrophysics_2006, storchi-bergmann_bipolar_2018, brum_close_2019, dallagnol_de_oliveira_gauging_2021}:

\begin{equation}
    \mathrm{M(H\beta)} = \frac{m_p\ \mathrm{L(H\beta)}}{n_e \alpha^\mathrm{eff}_\mathrm{H\beta}(\mathrm{T}) h \nu_\mathrm{H\beta}} 
\end{equation}

In this relation, $m_p$ is the proton mass, $\alpha^\mathrm{eff}_\mathrm{H\beta} = 3.02 \times 10^{-14}$\,cm$^3$\,s$^{-1}$ is the effective recombination coefficient of \hb\ at $10^4$\,K, assuming $n_e$\,=\,10$^2$\,cm$^{-3}$ and a recombination case B, and the term $h\nu_\mathrm{H\beta}$ is the \hb\ photon energy.

Since this relation is proportional to the line luminosity, the mass distributions are very similar to those of the LH$\beta$ in the bottom panels of Figures\,\ref{fig:hist_L_comp_lowL} and \ref{fig:hist_L_comp_highL}: the median  M(\hb) value is $3.26 \times 10^7$\,\msun\ for type 1 QSOs and $1.60 \times 10^7$\,\msun\ for the QSOs 2. We note that the first is $\sim 2$\,times larger than the latter. Regarding the mass contribution per NLR component, we found that $\sim$48\% of the ionized gas mass from the QSOs 1 NLR comes from the narrow component, while for QSOs 2 this fraction is $\sim$50\%. The median mass values are presented in Table\,\ref{tab:median_exc_lowL}.

As caveat, we note that the ionized gas mass can also be estimated from L(\oiii) \citep[e.g.][]{carniani_ionised_2015, trindade_falcao_hubble_2021, gatto_extent_2024}. In this case, the mass values are $\approx$50\% lower than those obtained from L(\hb) and are proportional to L(\oiii). They will thus be somewhat higher for QSOs 2 than for QSOs 1, with similar distribution as that for L(\oiii) in Figures\,\ref{fig:hist_L_comp_lowL} and \ref{fig:hist_L_comp_highL}. We tentatively attribute the difference in masses obtained from the two indicators to the different excitations of the gas, in which shocks decrease L(\oiii) relative to L(\hb) in type 1 QSOs, while the dominance of radiative excitation increases L(\oiii)/L(\hb) in QSOs 2.

\section{Results and Analysis - High Luminosity} \label{sec:res_highL}

\begin{table}
    \centering
    \caption{Median values for each QSO type for the luminosities, line ratios and electronic temperature parameters (discussed in Section\,\ref{sec:res_highL}) for the HL sample. The median values are shown in the following histograms as dashed lines -- for QSOs 1 -- and dotted lines -- for QSOs 2. Units are as follow: luminosities are in \lsun, masses are in \msun, and T$_\mathrm{e}$ is given in K. We also include the KS test p-value for each parameter.}
    \begin{tabular}{lccc}
\hline
\hline
Parameter & QSO 1 & QSO 2 & p$_\mathrm{KS}$ \\
\hline
log(L\oiii$\lambda$5007),n & 8.95 & 8.97 & 3.08$ \times 10^{-1}$ \\
log(L\oiii$\lambda$5007),b & 8.79 & 8.76 & 1.27$ \times 10^{-1}$ \\
log(L\oiii$\lambda$5007) & 9.18 & 9.17 & 6.37$ \times 10^{-2}$ \\
log(L \hb),n & 8.10 & 7.97 & 3.74$ \times 10^{-9}$ \\
log(L \hb),b & 8.12 & 7.75 & 1.77$ \times 10^{-63}$ \\
log(L \hb) & 8.24 & 8.15 & 3.74$ \times 10^{-9}$ \\
\oiii$\lambda$5007 n/b & 1.54 & 1.76 & 8.46$ \times 10^{-2}$ \\
\hb\,NLR n/b & 1.02 & 1.63 & 8.63$ \times 10^{-19}$ \\
log(L\oii) & 8.46 & 8.50 & 1.60 $ \times 10^{-1}$ \\
log(L\neiii) & 8.28 & 8.03 & 6.58$ \times 10^{-31}$ \\
log(L\oiii$\lambda$4363) & 8.03 & 7.50 & 3.87$ \times 10^{-53}$ \\
log(L\nev) & 8.37 & 7.85 & 8.31$ \times 10^{-110}$ \\
\nev/\neiii & 1.17 & 0.61 & 3.11$ \times 10^{-71}$ \\
\neiii/\oii & 0.64 & 0.36 & 7.75$ \times 10^{-43}$ \\
\oiii$\lambda$5007/\oii & 5.69 & 5.13 & 1.34$ \times 10^{-3}$ \\
\oiii$\lambda$5007/\hb & 9.18 & 11.00 & 9.81$ \times 10^{-13}$ \\
\oiii$\lambda$5007/\hb,n & 7.39 & 10.75 & 1.38$ \times 10^{-29}$ \\
\oiii$\lambda$5007/\hb,b & 6.25 & 10.78 & 2.29$ \times 10^{-45}$ \\
T$_\mathrm{e}$ & 20913 & 11985 & 5.47$ \times 10^{-41}$ \\
log(M \hb) & 7.66 & 7.57 & 3.74$ \times 10^{-9}$ \\
\hline
\end{tabular}
    \label{tab:median_exc_highL}
\end{table}

\subsection{Emission line luminosities}

\begin{figure*}
    \centering
    \includegraphics[width=0.85\textwidth]{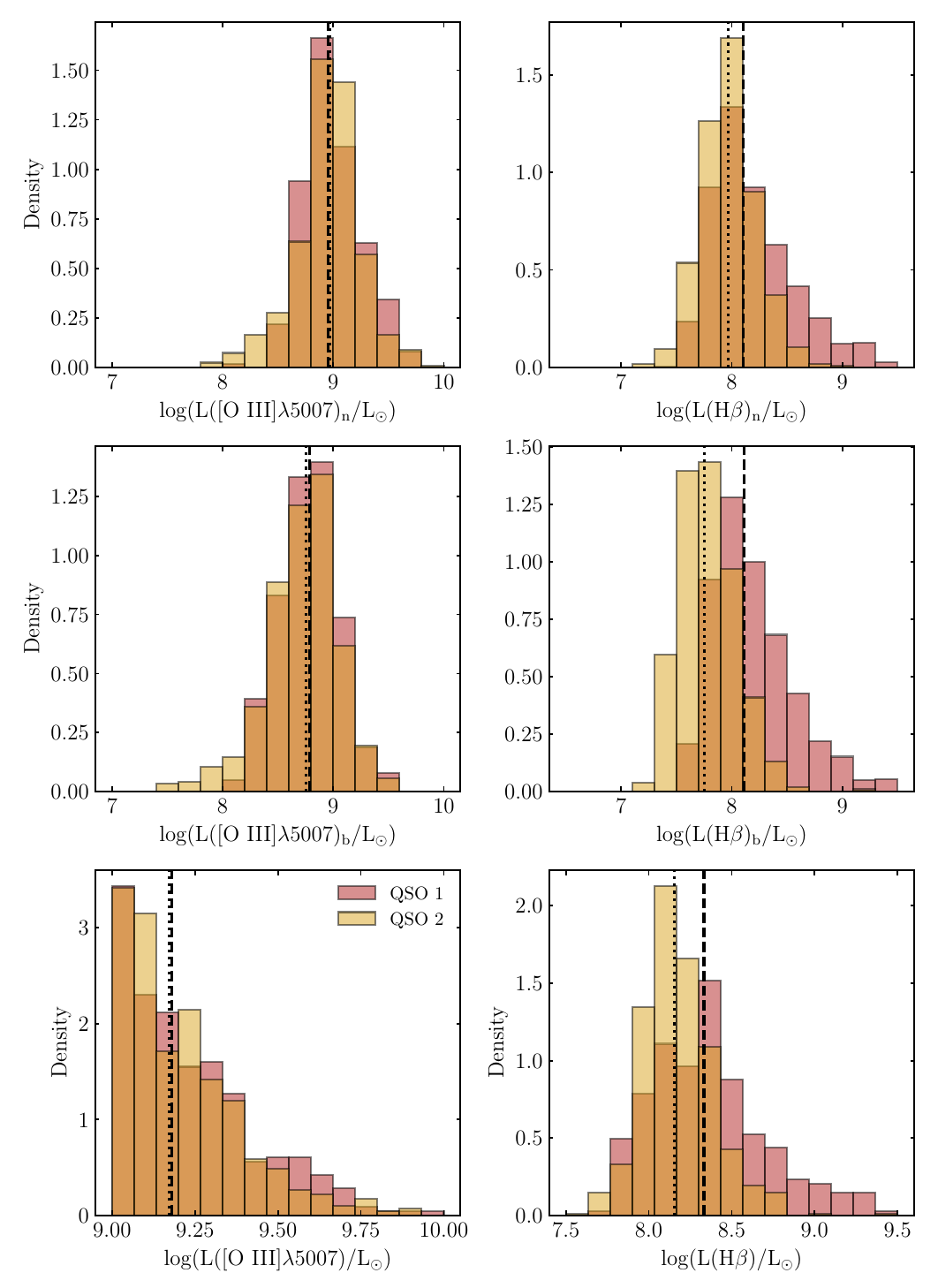}
    \caption{From top to bottom: normalized NLR$_\textrm{n}$, NLR$_\textrm{b}$ and total NLR luminosity distributions for the \oiii$\lambda$5007 and \hb\ emission lines for the HL sample. The QSO 1 and QSO 2 populations are shown as red and yellow histograms, respectively, showing the median of each population as black dashed or dotted vertical black lines.}
    \label{fig:hist_L_comp_highL}
\end{figure*}

Similarly to Section\,\ref{sec:res_lowL}, where we presented the results for the LL sample, we now present and discuss the results for the HL sample. The median values of the distributions as well as the corresponding KS-test p-values (p$_\mathrm{KS}$) are shown in Table\,\ref{tab:median_exc_highL}.  Although the differences in the sizes of the QSO 1 and 2 samples were smaller than for the LL sample, we have also performed the same random resampling test as done for the LL sample, only this time matching the larger QSO 2 sample size to that of the smaller QSO 1 sample. The same results regarding the median and KS-test have been found also for the HL sample, namely, the results are not affected by the differences in the number of type 1 and 2 sources.

\medskip

\noindent{\bf \oiii\ and \hb\ --} The panels of Figure\,\ref{fig:hist_L_comp_highL} show that the component-wise \oiii$\lambda$5007 luminosity distribution for the HL sample does not show bimodality for any of the NLR components. This is also seen for the full profile luminosity, with the number density of objects declining for higher luminosities for both QSO types. The medians for the component-wise and the entire emission profile are similar between both types, with all distributions presenting significant p$_\mathrm{KS}$ values larger than 0.05, which also indicates the similarities of the \oiii$\lambda$5007 line at the highest luminosities. Nevertheless, as discussed for the LL sample, the peak of the QSOs 2 L\oiii\ distribution seems not to be probed by the HL sample, and does not allow the comparison between the peak of the two distributions, that seems to occurs in the LL range of L\oiii.

For \hb, the distributions are different, luminosities extending to higher values for QSOs 1, but with a similar overlap with QSOs 2 to the one found for the LL sample. The component-wise and full profile distribution present typically larger values for type 1 QSOs, due to skewness toward the highest values found in all histograms.

\begin{figure*}
    \centering
    \includegraphics[width=0.85\textwidth]{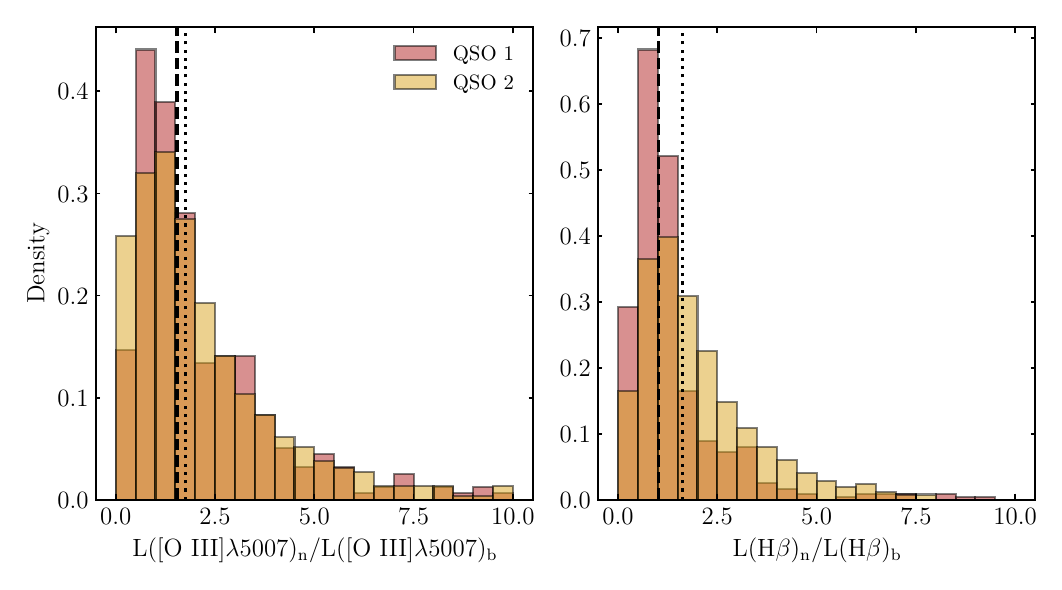}
    \caption{Normalized \oiii$\lambda$5007 NLR$_\mathrm{n}$/NLR$_\mathrm{b}$ (left) and \hb\ NLR$_\mathrm{n}$/NLR$_\mathrm{b}$ components ratios (right) for the HL sample. The QSO 1 and QSO 2 populations are shown as red and yellow histograms, respectively, showing the median of each population as black dashed or dotted vertical black lines.}
    \label{fig:hist_na_br_highL}
\end{figure*}

The luminosity ratio between the NLR$_\mathrm{n}$ and NLR$_\mathrm{b}$ components for \oiii$\lambda$5007 and \hb\ are shown in the panels of Figure\,\ref{fig:hist_na_br_highL}. For \oiii, the distribution is similar between both QSO types, as stated by the high p$_\mathrm{KS}$ value. The medians are, respectively, 1.54 and 1.76, which reveal that the majority of the mass comes from the narrow component. For \hb, the distribution is skewed toward larger values for type 2 QSOs than for type 1's, being visually distinct and presenting p$_\mathrm{KS} \ll 0.05$. The medians for type 1 and 2 QSOs are, respectively, 1.02 and 1.63, which reveals an excess of mass in the narrow components of type 2 QSOs, while roughly the same amount of mass in the two regions is typically found for type 1's.

\begin{figure*}
    \centering
    \includegraphics[width=0.85\textwidth]{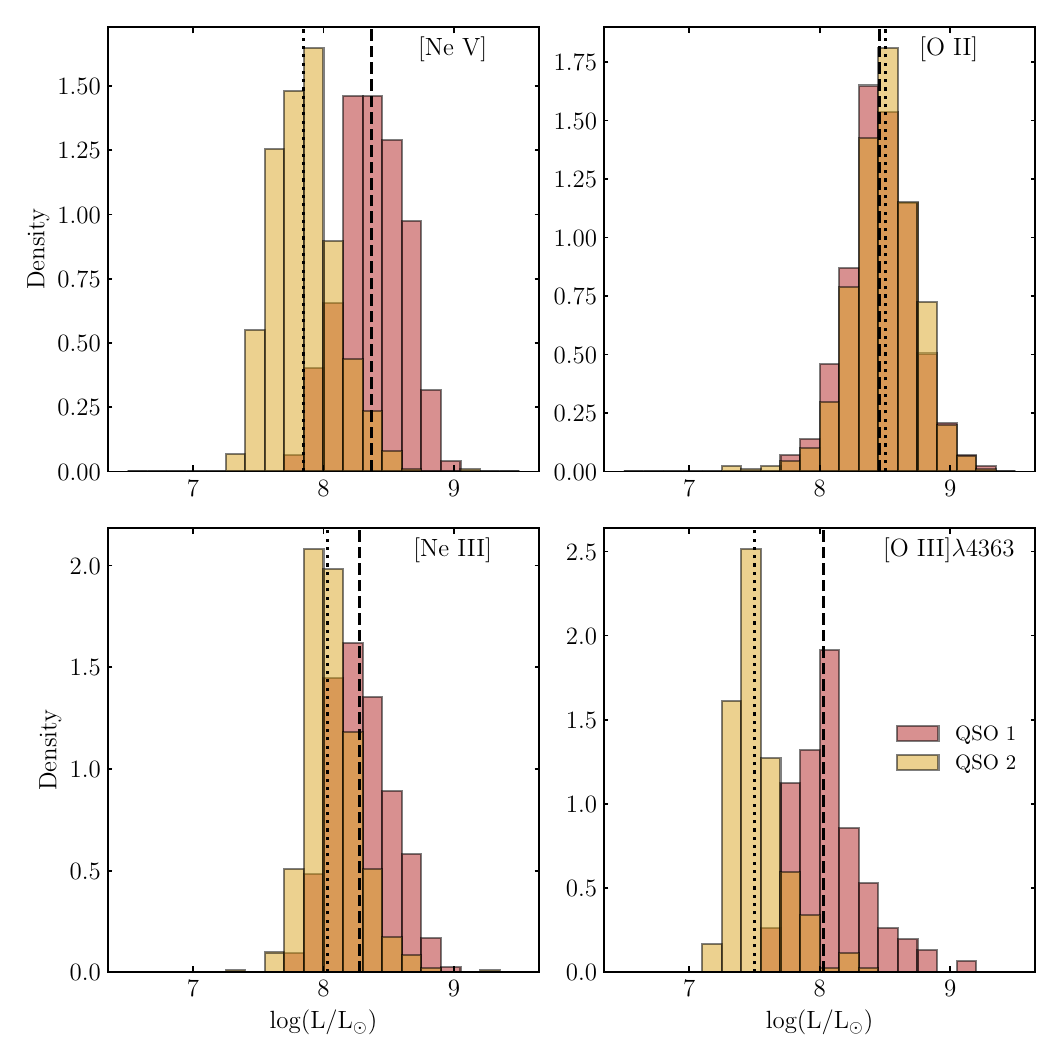}
    \caption{Normalized luminosity distributions of the \nev, \oii, \neiii\ and \oiii$\lambda$4363 emission lines for the HL sample for the QSO 1 and QSO 2 populations, shown as red and yellow histograms, respectively, with the corresponding median shown as black dashed and dotted vertical black lines.}
    \label{fig:hist_lum_highL}
\end{figure*}

\medskip
\noindent{\bf \nev, \oii, \neiii\ and \oiii$\lambda$4363 --} The luminosity distributions for these lines are shown in the four panels of Figure\,\ref{fig:hist_lum_highL}, for both QSO types. We find a clear bimodality for both \nev\ and \oiii$\lambda$4363 emission lines, in which type 1 QSOs are found to be more luminous than type 2's. For \neiii, a weaker bimodality is found, but QSOs 1 also present higher luminosities than QSOs 2. For \oii, no bimodality is found and the luminosity distributions are very similar, as confirmed by the high p$_\mathrm{KS}$ value.

Similarly to the lower luminosity sample, the bimodalities of these higher excitation lines indicate we are seeing higher radiative excitation regions in the type 1s with respect to the type 2s. 

\subsection{Emission-line ratios}
\label{sec:exc_res_highL}
\begin{figure*}
    \centering
    \includegraphics[width=0.85\textwidth]{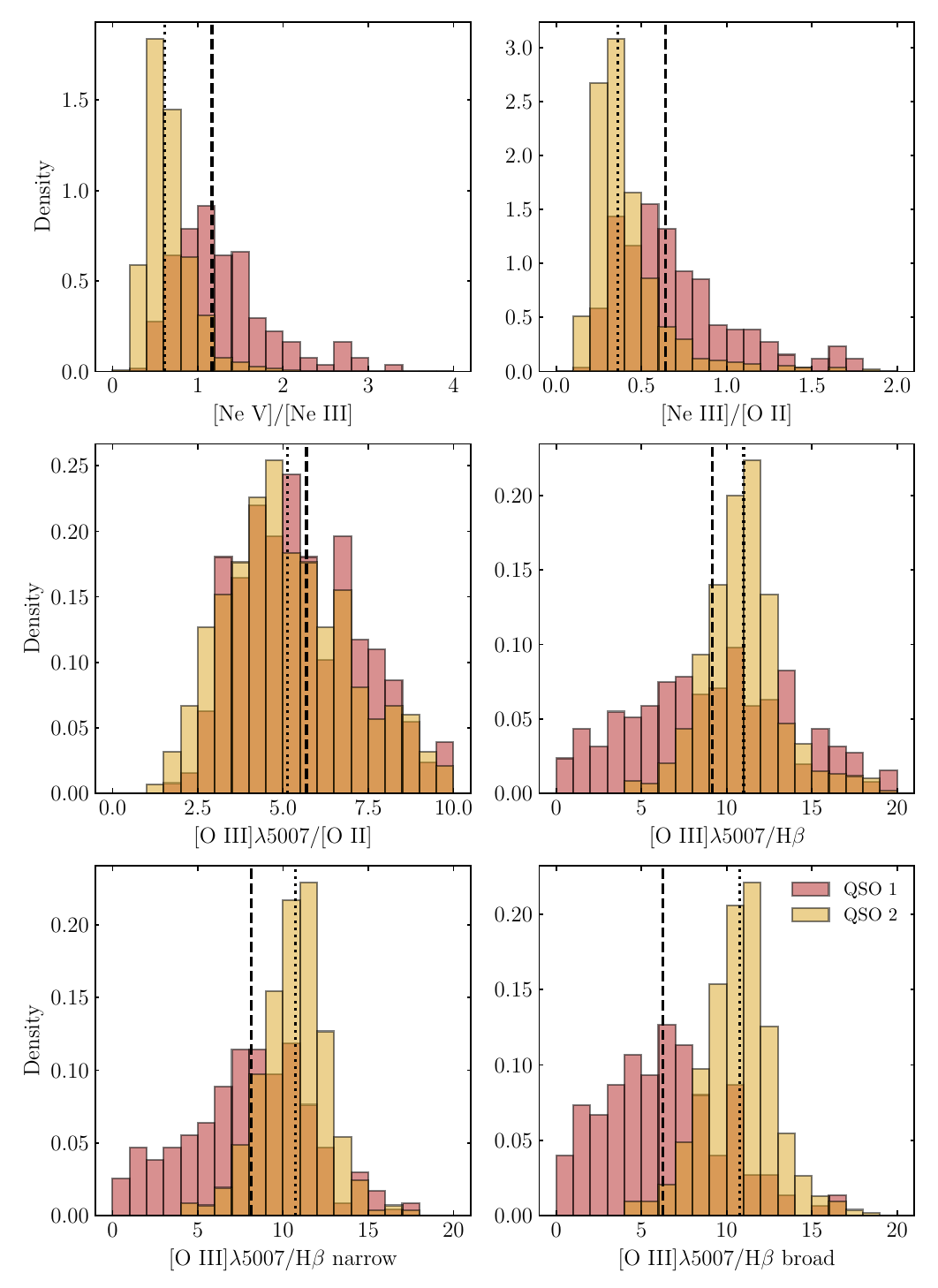}
    \caption{Normalized distributions of the \nev/\neiii, \neiii/\oii, \oiii$\lambda$5007/\oii\ and \oiii$\lambda$5007/\hb\ line ratios for the high-luminosity sample, also including the individual NLR components ratios for the latter. The distributions are for the QSO 1 and QSO 2 populations, shown as red and yellow histograms, respectively. The median of each population are shown as black dashed or dotted vertical black lines.}
    \label{fig:hist_ratios_highL}
\end{figure*}

Similar to Section \ref{sec:exc_res}, we present here the results and discussion regarding the emission-line ratios for the high-luminosity sample, shown in Figure \ref{fig:hist_ratios_highL}.

\medskip
\noindent{\bf \nev/\neiii\ -- } This ratio shows higher values for type 1 objects than for type 2, similarly to the LL sample, consistent with the scenario in which the inner regions of the central engine of the AGN is seen more directly for these objects, in relation to type 2s, or due to evolution from an obscure phase as a type 2 to a clearer view of the AGN as a type 1 after the blowout phase in the evolutionary scenario.

\medskip
\noindent{\bf \neiii/\oii\ -- } This ratio reaches higher values for QSOs 1 than for QSOs 2. Since the distribution is similar to the LL sample, we interpret the results the same way (see Section\,\ref{sec:exc_res}), also corresponding to the scenarios above.

\medskip
\noindent{\bf \oiii$\lambda$5007/\oii\ -- } The distributions of the values of this ratio is visually similar for the two QSO types. However, the KS-test reproves the statistical similarity between both distributions, since p$_\mathrm{KS} < 0.05$. The median values for QSOs 1 are higher than for type 2s, and both are higher than for the LL sample. The latter part can be explained due to the higher \oiii\ luminosity selection criteria, considering that the \oii\ has a larger possible contribution from stellar ionisation.

\medskip 
\noindent{\bf \oiii$\lambda$5007/\hb\ -- } This ratio presents a clear bimodality, being more concentrated and presenting a larger median value for QSOs 2 with respect to QSOs 1. This trend is also seen for the individual NLR components, shown in the bottom of Fig. \ref{fig:hist_ratios_highL}. Similarly to the low-luminosity sample, while there are essentially no ratios below 5 in QSOs 2 (5 objects), there are many more (52 objects) in QSOs 1, which still supports the prevalence of shocks in the QSOs 1 relative to QSOs 2, as discussed in the low-luminosity case. But other than that, the distribution of this ratio for the QSOs 1 is distinct from the ones observed for the LL sample, for which the \oiii$\lambda$5007/\hb\ were skewed towards lower values. For this higher luminosity sample, the line ratios extend to higher values, showing a larger overlap with the QSOs 2 high values.

\begin{figure}
    \centering
    \includegraphics[width=0.96\linewidth]{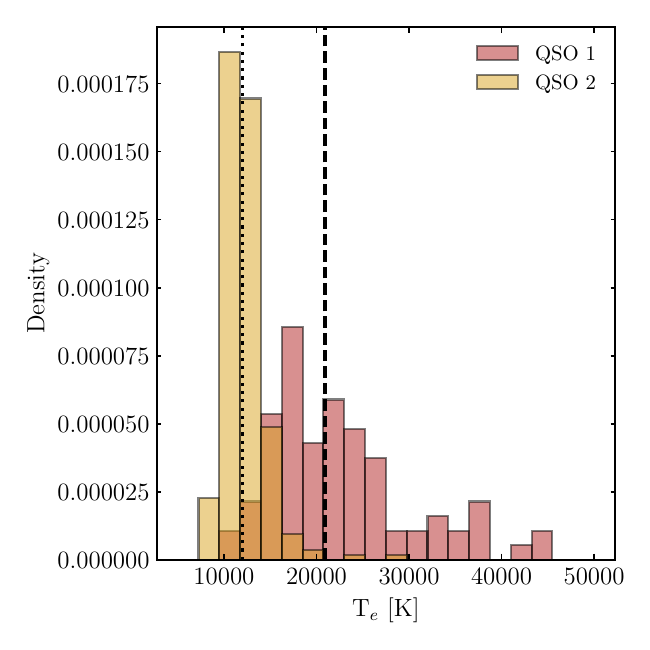}
    \caption{Normalized electronic temperature for the high-luminosity sample for the QSO 1 and QSO 2 populations, presented as red and yellow histograms, respectively, showing the median of each population as black dashed or dotted vertical black lines.}
    \label{fig:te_highL}
\end{figure}

\subsection{Electronic temperature} \label{sec:te_res_highL}

For the high-luminosity sample, the SNR in the \oiii$\lambda$4363 line was high enough to allow the calculation of the electronic temperature. We present the calculated T$_\mathrm{e}$ distributions in Figure\,\ref{fig:te_highL}, which reveals the bimodality between QSOs 1 and QSOs 2: the first possess higher T$_\mathrm{e}$ than the latter, with medians centered around $21 \times 10^3$\,K and $12 \times 10^3$\,K, respectively. 

While the QSOs 2 have median temperatures that are characteristic of gas excited by radiation, the QSOs 1 reach temperature values as high as $40 \times 10^3$\,K, much higher than expected from purely radiative excitation, showing evidence of the presence of shocks  \citep[e.g.][]{dopita_spectral_1995, dopita_spectral_1996, osterbrock_astrophysics_2006, allen_mappings_2008}. This shock excitation scenario is also supported by the lower \oiii$\lambda$5007/\hb\ ratios found for type 1 QSOs in relation to the larger values of the type 2 sample. This can be due to two possible scenarios: a different orientation of the nuclear source, in the framework of the Unified Model, or due to different properties of the NLR of the two samples.

Considering the Unified Model scenario, in which we are looking more directly and thus reaching gas closer to the ionizing source in QSOs 1 than in QSOs 2: part of the highest ionised clouds (e.g. producing \nev) are in the hottest regions that are hidden by the obscuring torus in the QSOs 2, which is also embedded within the magnetic field produced by the rapidly rotating plasma that can lead to shocks in the ionised gas \citep[e.g.][]{begelman_theory_1984, allen_mappings_2008}. Therefore, due to a visibility effect, the shock signature can be seen more easily in QSO 1s. 

\begin{figure}
    \centering
    \includegraphics[width=0.96\columnwidth]{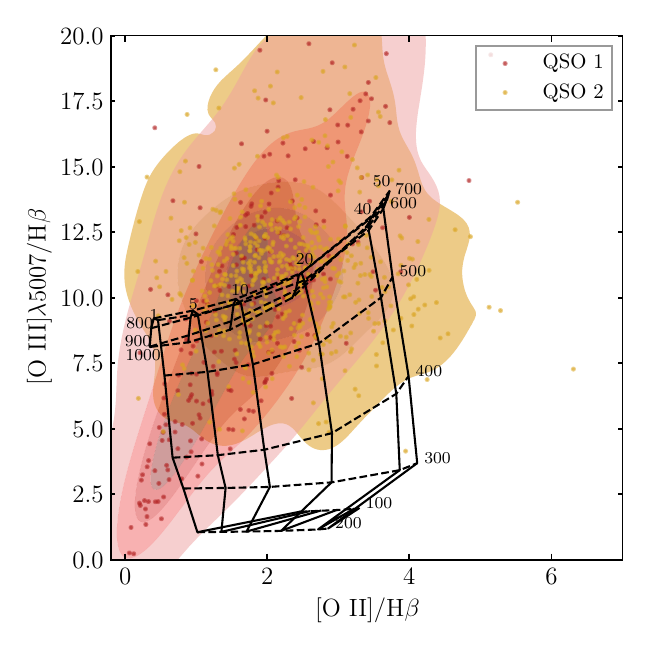}
    \caption{\oiii$\lambda$5007/\hb\ versus \oii/\hb\ diagram and shock models. The QSO 1 and QSO 2 from the HL sample are shown as red and yellow dots, respectively. In black, we overplot the \citet{allen_mappings_2008} shock+precursor models for solar metallicity and n$_\mathrm{e}=100$\,cm$^{-3}$. As continuous lines, we show the magnetic field (B) with varying strengths of 1, 5, 10, 20, 40 and 50\,$\mu$G; as dashed lines, we show the different velocities (v) for the model, from 100 to 1000\,\kms in steps of 100\,\kms. The values of B and v are shown at the edges of the grid.  We also include the density contours of the distributions of the points for both QSO 1 and 2 samples, as a red and yellow shaded regions, respectively}.
    \label{fig:shock_highL}
\end{figure}

 If we consider the evolutionary scenario \citep[e.g.][]{araujo_nuclear_2023, pierce_galaxy_2023}, mergers or interactions disrupt the gravitational potential of the host galaxy, driving gas inwards which can fuel stellar formation and subsequently feed the SMBH. The galaxy then becomes active, with a central engine buried in gas and dust, which obscures the accretion disk -- as a type 2 source. The radiation field eventually ionises and expels the surrounding gas and dust, revealing the inner regions of the AGN, that becomes, then, a type 1 object \citep{hopkins_unified_2006, buchner_obscuration-dependent_2015}. 

We test the hypothesis of the presence of shocks in the NLR of type 1 QSOs by overplotting the \citet{allen_mappings_2008} shock+precursor models over a \oiii$\lambda$5007/\hb\ versus \oii/\hb\ diagram, as shown in Figure\,\ref{fig:shock_highL}, using their solar chemical abundance models with n$_\mathrm{e}=100$\,cm$^{-3}$. The grid consists of black, continuous lines denoting equal magnetic field strengths with values of 1, 5, 10, 20, 40 and 50\,$\mu$G; and black, dashed lines indicating the shock velocities, ranging from 100 to 1000\,\kms, with steps of 100\,\kms.  We find that our data points for the type 1 QSOs, shown in shades of red, are better covered by the shock models (in particular if we consider the possibility of some reddening affecting the \oii/H$\beta$ ratio, which, corrected, would give somewhat higher ratios). The bulk of the QSOs 2, on the other hand, falls mostly above the grid, although with a large scatter. This result, along with the higher electronic temperature, supports the dominance of shocks in the QSOs 1, while the higher values of both  \oiii/\hb\ and \oii/\hb\ ratios for QSOs 2 are likely to be produced by radiative excitation. 

\subsection{Ionised gas mass}

We calculate the ionised gas mass using the same approach previously presented for the LL sample. We obtain a median mass of $4.55 \times 10^7$\,\msun\ for type 1 QSOs and $3.73 \times 10^7$\,\msun\ for QSOs 2. The results are presented in \ref{tab:median_exc_highL}, where we also indicate the p$_\mathrm{KS}$ value for the mass distributions, which is smaller than 0.05. We again opted not to include a separate histogram of the mass distributions since its derivation is directly proportional to L(\hb), and therefore it would be similar to the bottom right panel of Figure\,\ref{fig:hist_L_comp_highL}.

As for the LL sample, the HL sample also shows a significant statistical difference between the ionised gas mass distributions, with the masses $\sim$20\% larger for type 1 QSO than for type 2 objects.

The caveat discussed for the LL sample also applies for the HL sample. For the HL sample, the ionized gas mass estimated from L\oiii\ \citep[e.g.][]{carniani_ionised_2015, trindade_falcao_hubble_2021, gatto_extent_2024} is also $\approx$50\% lower than those obtained from LH$\beta$ and are proportional to L\oiii. They will thus be similar for QSOs 2 and QSOs 1, with similar distribution as that for L\oiii\ in Figure\,\ref{fig:hist_lum_highL}. As in the case of the LL sample, we attribute the difference in masses obtained from the two indicators to the different excitation mechanisms of the gas.

We also investigate the mass fraction per NLR component: we find that $\sim$73\% of the \hb\ mass originates from the narrow component of type 1 QSOs, while for type 2 QSOs this fraction is smaller, comprising $\sim$65\% of the entire mass. This is very different from the result found for the lower luminosity sample, in which both median fractions were around 50\%. 

As a caveat, we note that these percentages seem to be at odds with the component-wise ratios presented in Table\,\ref{tab:median_exc_highL}, due to the fact that for the calculation of the percentages above, we take into account the cases in which only one NLR component is reliably detected (either narrow or broad). These cases skew the medians towards higher values.

\section{Conclusions}

In this work, we have analysed the gas excitation properties of the Narrow-Line Region (NLR) of a spectroscopic sample of 2301 QSOs at redshifts $0.4 \leq z \leq 0.5$, comparing QSOs 1 and 2. We further subdivide this into 2 subsamples: a low luminosity (LL) subsample, with objects with L$_{\mathrm{\oiii}}< 10^9$\lsun; and a high luminosity (HL) subsample, for objects with L$_{\mathrm{\oiii}}> 10^9$\lsun. We present the methodology of the analysis of the AGN continuum and emission lines, that include two components for the NLR \oiii\ and \hb\ emission lines: a narrow and a broad. We find that there are differences between QSOs 1 and 2 regarding their line luminosities and excitation, as follows:

\begin{itemize}

    \item The L(\oiii$\lambda$5007) of the broad component is, on average, 50\% higher for QSOs 2 than for QSOs 1 for the LL sample. This  could be explained with an excess of more kinematically disturbed gas in type 2 QSOs. For the HL sample, the \oiii\ luminosity distribution is similar for the two types, which suggests the dominance of radiative excitation and similar kinematic disturbances in both QSO types at the highest luminosities; 

    \item Electronic temperatures could be measured for the HL sample, and are are higher for QSOs 1 (median T$_\mathrm{e} = 21 \times 10^3$\,K) than for QSOs 2 (median T$_\mathrm{e} = 12 \times 10^3$\,K). This supports the presence of shocks as an additional excitation mechanism in the QSOs 1, while for QSOs 2 the temperature is consistent with excitation via radiation only;

    \item The presence of shocks in the NLR of QSOs 1 is also supported by the emission-line ratios: QSOs 1 show higher \nev/\neiii\ but lower \oiii$\lambda$5007/\hb\ ratios than in QSOs 2, with the latter comprising values typical of LINERs. The QSOs 2 show values higher that peak in the interval 10--20, as expected for Seyfert galaxies; 
    
    \item The ionised gas masses have been obtained via the \hb\ luminosity as an emission-line tracer. For the LL sample, the median values are $3.26 \times 10^7$\,\msun\ for the QSOs 1 and $1.60 \times 10^7$\,\msun\ for the QSOs 2, while for the HL sample, they are, respectively, $4.55 \times 10^{7}$\,\msun\ and $3.73 \times 10^{7}$\,\msun. As a caveat, \oiii\ can also be used to determine the ionised gas mass, but provides $\approx$50\% of the mass obtained with \hb, with QSOs 2 presenting a larger ionised gas mass then type 1 objects in general. We interpret this as another indicator of the presence of distinct excitation mechanisms of the gas in QSOs 1 and 2;
    
    \item When comparing the LL and HL samples, there is an excess of type 1's in the LL sample and of type 2's in the HL sample, revealing that L(\oiii) is overall higher in type 2's.

\end{itemize}

The higher excitation, the signature of shocks and higher temperature (for the HL sample) seen in QSOs 1 when compared to QSOs 2 are in principle consistent with the Unified Model, in which we are seeing more of these regions in type 1 sources due to the more face-on orientation. However, we cannot exclude the possibility that this could be due to evolution from type 2's to type 1's. In the evolutionary scenario, luminous AGN are triggered by galaxy mergers and interactions and are initially in an obscured type 2 phase that later evolves to a type 1 AGN after the clearing of the obscuring gas by the AGN feedback. Support for the evolutionary scenario is also given by the higher number of type 2 sources in the HL sample and higher number of type 1 sources in the LL sample, revealing that the peak of the L\oiii\ distribution for the QSOs 1 occurs in the LL (low-luminosity) range, while that of the QSOs 2 occur at higher values, of $\approx$ 10$^9$\,L$_\odot$. This reveals that type 2 sources, on average, host a more powerful AGN that, in its evolution, end up clearing the excess dust and gas and becoming a lower-luminosity type 1 AGN.

A follow up work investigating the NLR gas kinematics is under preparation, which aims to reveal the connection between the AGN excitation source and the NLR kinematics, looking for AGN outflows and calculating their powers.

\section*{Acknowledgements}

The authors would like to thank the anonymous referee for their valuable suggestions and insights. The authors would also like to thank the brazilian Conselho Nacional de Desenvolvimento Científico e Tecnológico (CNPq) for the support for this research. RR acknowledges support from  Conselho Nacional de Desenvolvimento Cient\'{i}fico e Tecnol\'ogico  (CNPq, Proj. 311223/2020-6, 304927/2017-1, 400352/2016-8, and 404238/2021-1), Funda\c{c}\~ao de Amparo \`{a} Pesquisa do Rio Grande do Sul (FAPERGS, Proj. 19/1750-2 and 24/2551-0001282-6) and Coordena\c{c}\~ao de Aperfei\c{c}oamento de Pessoal de N\'{i}vel Superior (CAPES, Proj. 0001). 

\section*{Data Availability}

The spectra used in this work are available for download at the SDSS-IV DR16 website (https://www.sdss4.org/dr16/). All figures produced from these data, and the data itself, can be shared on reasonable request to the corresponding author.



\bibliographystyle{mnras}
\bibliography{references} 





\bsp	
\label{lastpage}
\end{document}